\documentclass{aa}     

\usepackage{graphicx}

\newcommand{\Mpc}{$h^{-1}$\thinspace Mpc}

\newcommand{\etal}{et al.}
\def\apj{ApJ}

\def\aa{A\&A}

\begin{document}


\title{Clusters and Superclusters in the Las Campanas
Redshift Survey}

\author {J. Einasto\inst{1}, M. Einasto\inst{1}, G. H\"utsi\inst{1}, E.
Saar\inst{1},  D. L. Tucker\inst{2}, E. Tago\inst{1}, V. M\"uller\inst{3},
 P. Hein\"am\"aki\inst{1,4}, S. S. Allam\inst{2,5} }

\authorrunning{J. Einasto et al.}

\offprints{J. Einasto }

\institute{Tartu Observatory, EE-61602 T\~oravere, Estonia
\and
 Fermi National Accelerator Laboratory, MS 127, PO Box 500, Batavia,
IL 60510, USA
\and
Astrophysical Institute Potsdam, An der Sternwarte 16,
D-14482 Potsdam, Germany
\and
Tuorla Observatory, V\"ais\"al\"antie 20, Piikki\"o, Finland
\and
 Dept. of Astronomy, New Mexico State University, Las Cruces, 
  NM 88003-8001, USA
}
\date{ Received   2003 / Accepted ...  }

\titlerunning{LCRS clusters and superclusters}

\abstract{ We use a 2-dimensional high-resolution density field of
galaxies of the Las Campanas Redshift Survey (LCRS) with a
smoothing length 0.8~\Mpc\ to extract clusters and groups of galaxies,
and a low-resolution field with a smoothing length 10~\Mpc\ to find
superclusters of galaxies.  We study the properties of these density field (DF)
clusters and superclusters, and compare the properties of the
DF-clusters and superclusters with those of Abell clusters and
superclusters and LCRS groups. We show that among the cluster
samples studied the DF-cluster sample best describes the large-scale
distribution of matter and the fine structure of superclusters.  We
calculate the DF-cluster luminosity function and find that clusters in
high-density environments are about ten times more luminous than those in
low-density environments.  We show that the DF-superclusters that
contain Abell clusters are richer and more luminous than the
DF-superclusters without Abell clusters.  The distribution of
DF-clusters and superclusters shows the hierarchy of systems in the
universe.

\keywords{cosmology: observations -- cosmology: large-scale structure
of the Universe; clusters of galaxies}
}

\maketitle

\section{Introduction}

The basic tasks of observational cosmology are to describe the
distribution of various objects in the universe and to understand the
formation and evolution of these structures.  One means for describing
the structure is the density field method.  In this method
the distribution of discrete objects (galaxies and clusters of
galaxies) is substituted by the density field calculated by smoothing
the discrete distribution.  This method has the advantage
that it is easy to take into account various selection effects which
distort the distribution of individual objects.  The density field can
be applied to calculate the gravitational field as done in the
pioneering study by \cite{1982ApJ...254..437D}, to investigate
topological properties of the universe (Gott et al. 
~\cite{gmd86}), and to map the universe and to find superclusters and voids
(Saunders et al. \cite{s91}, Marinoni et al. \cite{mar99}, Hoyle et
al. \cite{h2002}, Basilakos et al. \cite{bpr01}).

In this paper we use the density field of galaxies to find clusters
and superclusters of galaxies. This method was introduced by Einasto
et al. (\cite{e03b}, hereafter Paper I) and applied to the Early Data
Release of the Sloan Digital Sky Survey.  Here we apply the density
field method to the Las Campanas Redshift Survey (LCRS).  The LCRS is
essentially a 2-dimensional survey; however, using the LCRS data we
obtain useful information for clusters and superclusters that is not
yet available from 3-dimensional surveys of comparable depth.  As in
Paper I we use the high-resolution density field to find clusters and
groups of galaxies as enhancements of the field, and the
low-resolution density field to construct a catalogue of superclusters
of galaxies.  For simplicity, we use the term ``DF-clusters'' for both
groups and clusters found from the high-resolution density field of
galaxies; similarly, we use the term ``DF-superclusters'' for large
overdensity regions detected in the low-resolution density field. In
identifying the DF-clusters and superclusters we take into account
known selection effects.  The main selection effect is due to the
limited range of apparent magnitudes used in redshift surveys.  We
assume that galaxy luminosities are distributed according to the
Schechter (\cite{Schechter76}) luminosity function, and find the
correction for galaxies with luminosities outside the observing window
applying the Schechter parameters as found by H\"utsi et
al. (\cite{hytsi02}, hereafter H03) for the LCRS.  We shall
investigate statistical properties of DF-clusters and superclusters,
and study the role of these clusters and superclusters as tracers of
the structure of the universe. We compare the distribution of
DF-clusters and superclusters with that of the LCRS loose groups
(Tucker et al \cite{Tucker00}, hereafter TUC), and of Abell clusters
and of superclusters traced by Abell clusters (Abell superclusters)
(Einasto et al. \cite{e2001}, hereafter E01).  This study is carried
out in the framework of preparation for the analysis of results of the
Planck mission to observe the Cosmic Microwave Background radiation.

In Sect. 2 we give an overview of observational data. In Sect. 3 we
identify the DF-clusters, discuss selection effects in the LCRS,
analyse properties of DF-clusters, and derive the luminosity function
of DF-clusters.  Similarly, in Sect. 4 we compose a catalogue of
DF-superclusters and analyse these systems as tracers of the structure
of the universe.  Sect. 5 brings our conclusions.  In Tables
~\ref{tab:sc1} and \ref{tab:sc2} we list the DF-superclusters and
their identification with conventional superclusters.  The
three-dimensional distribution of clusters and superclusters, as well
as colour versions of the figures with density field maps, are
available on the Tartu Observatory website
(www.aai.ee/$\sim$maret/cosmoweb.htm).

\section{Observational data}

\subsection{LCRS galaxies and loose groups}

The LCRS (Shectman et al. \cite{Shec96} ) is an optically selected
galaxy redshift survey that extends to a redshift of 0.2 and covers
six $1.5 \times 80$ degree slices containing a total of $23,697$
galaxies with redshifts.  Three slices are located in the Northern
Galactic cap centred at the declinations $\delta= -3^{\circ},~
-6^{\circ},~ -12^{\circ}$, and three slices are located in the
Southern Galactic cap centred at the declinations $\delta=
-39^{\circ},~ -42^{\circ},~ -45^{\circ}$.  The thickness of the survey
slices at the mean redshift of the survey ($z \approx 0.1$) is
approximately $7.5$~\Mpc.  Throughout this paper, the Hubble constant
$h$ is expressed in units of $100$ km~s$^{-1}$~Mpc$^{-1}$.

The spectroscopy of the survey was carried out via a 50 or a 112 fibre
multi-object spectrograph; therefore the selection criteria varied
from field to field. The nominal apparent magnitude limits for the 50
fibre fields were $m_1=16.0 \le R \le m_2=17.3$, and for the 112 fibre
fields $m_1=15 \le R \le m_2=17.7$.  The general properties of the 50
fibre and the 112 fibre groups agree well with group properties found
from other surveys.  We note that in the case of one slice, $\delta=
-6^{\circ}$, all observations were carried out with the 50-fibre
spectrograph only.  On the basis of the LCRS galaxies TUC extracted a
catalogue of loose groups of galaxies; a group had to contain at least
3 galaxies to be included in the catalogue (for more details on the
compilation of the group catalogue see TUC). Data on the LCRS slices
are given in Table~1: $RA$ -- the mean right ascension of the slice,
$\Delta RA$ -- the width of the slice (both in degrees), $N_{gal}$ --
the number of galaxies, $N_{DF}$ -- the number of DF-clusters,
$N_{LC}$ -- the number of loose groups by TUC, $N_{A}$ -- the number
of Abell clusters, and $N_{scl}$ -- the number of DF-superclusters.

{\footnotesize
\begin{table}
      \caption[]{Data on LCRS galaxies, clusters and superclusters}
         \label{Tab1}
      \[
         \begin{tabular}{rrrrrrrr}
            \hline
            \noalign{\smallskip}
            Slice $\delta$ & $RA$ & $\Delta RA$ 
	    & $N_{gal}$ &  $N_{DF}$
            & $N_{LG}$ & $N_{A}$&  $N_{scl}$ \\
            \noalign{\smallskip}
            \hline
            \noalign{\smallskip}
$-3^{\circ}$ & 191.4 & 81.0  &4065  & 1203  &289  & 18 & 19 \\
$-6^{\circ}$ & 189.8 & 77.9  &2323  &  952  &147 & 13 &  17 \\
$-12^{\circ}$& 191.4 & 81.1  &4482  & 1266  &276 & 11 &  15 \\
$-39^{\circ}$&  12.1 &113.8  &3922  & 1285  &256 &  28 & 18 \\
$-42^{\circ}$&  12.2 &112.5  &4158  & 1216  &265 &  19 & 14  \\
$-45^{\circ}$&  12.3 &114.1  &3753  & 1182  &263 &  20 & 17  \\

            \noalign{\smallskip}
            \hline
         \end{tabular}
      \]
   \end{table}
}

\subsection{Abell clusters and superclusters}

We shall use the catalogue of rich clusters of galaxies by Abell
(\cite{abell}) and Abell et al. (\cite{aco}) (hereafter Abell
clusters).  All published galaxy redshifts toward galaxy clusters, as
well as other data were collected by Andernach \& Tago (\cite{at98}).
>From that compilation we included in our study Abell clusters of all
richness classes (but excluded clusters from ACO's supplementary list
of S-clusters) with redshifts up to $z=0.13$.  The sample contains
1665 clusters, 1071 of which have measured redshifts for at least two
galaxies.  This sample was described in detail in E01, where an
updated supercluster catalogue of Abell clusters was presented. These
E01 superclusters were identified using the friend-of-friends
algorithm, first employed in studies of large-scale structure by
Turner \& Gott (\cite{tg76}) and Zeldovich et al. (\cite{zes82}).  All
clusters in a supercluster have at least one neighbour at a distance
not exceeding the neighbourhood radius of 24~\Mpc.

In the present paper we use the E01 catalogue as a reference to
identify density field superclusters with conventional ones. In Table
\ref{Tab1} and Fig. \ref{fig:2} we have used an updated version
(January 2003) of the compilation of redshifts of Abell clusters by
Andernach and Tago.

\begin{figure*}[ht]
\centering
\resizebox{0.45\textwidth}{!}{\includegraphics*{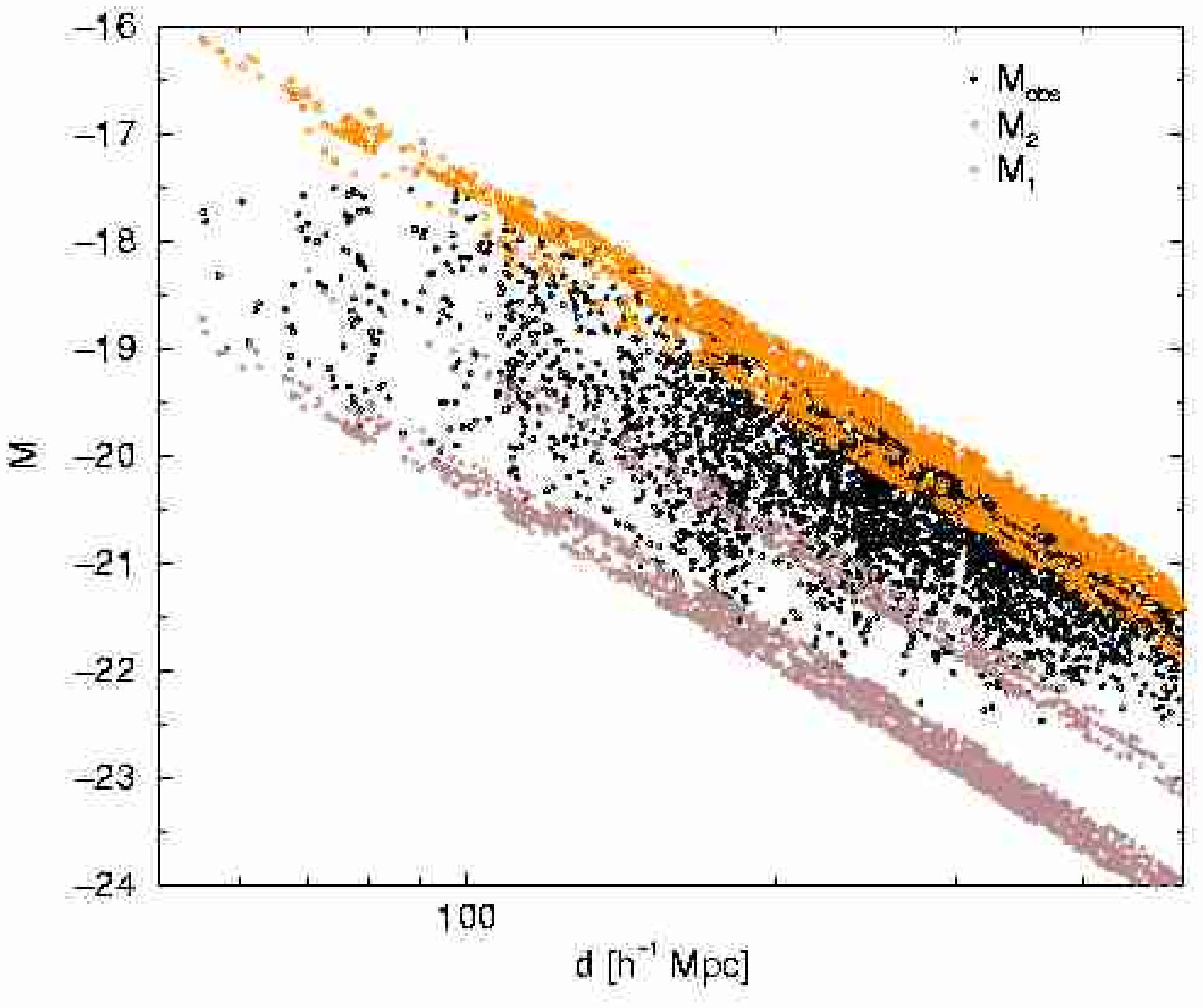}}\hspace{2mm}
\resizebox{0.45\textwidth}{!}{\includegraphics*{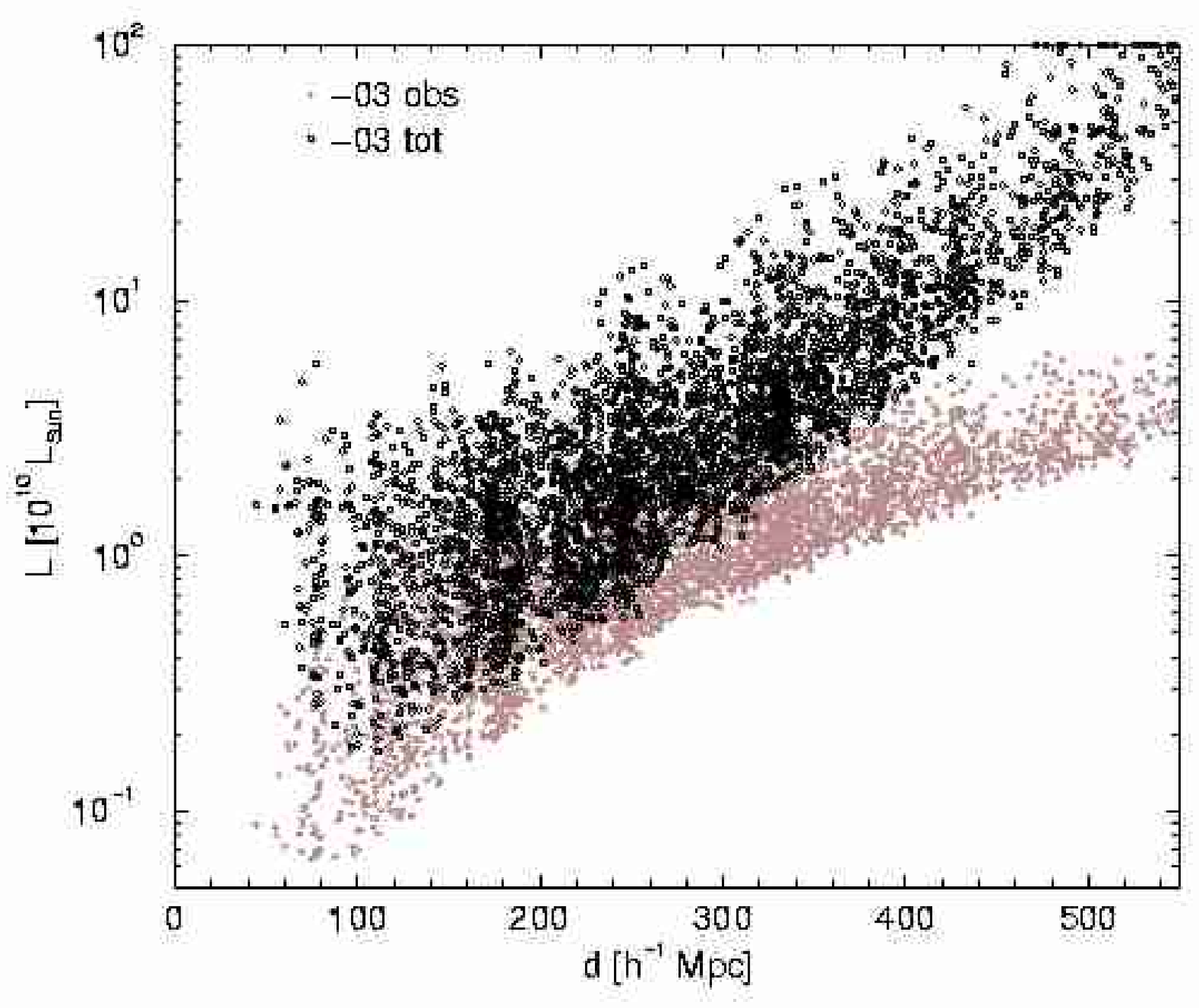}}\hspace{2mm}
\caption{The left panel shows the absolute magnitudes of galaxies, as well
as magnitudes of the luminosity window, $M_1$ and $M_2$, for the
$-3^{\circ}$ slice. The right panel gives the luminosities (weights)
of galaxies as a function of distance for the same slice.  In the left
panel black symbols mark the absolute magnitudes of observed galaxies,
the upper and lower strips with grey symbols show the absolute
magnitude limit $M_1$ and $M_2$. In the right panel grey symbols show
the observed luminosities of galaxies, black symbols are for total
luminosities corrected for the unobserved part of the luminosity
range.}
\label{fig:1}
\end{figure*}

\begin{figure*}[ht!]
\centering
\resizebox{0.45\textwidth}{!}{\includegraphics*{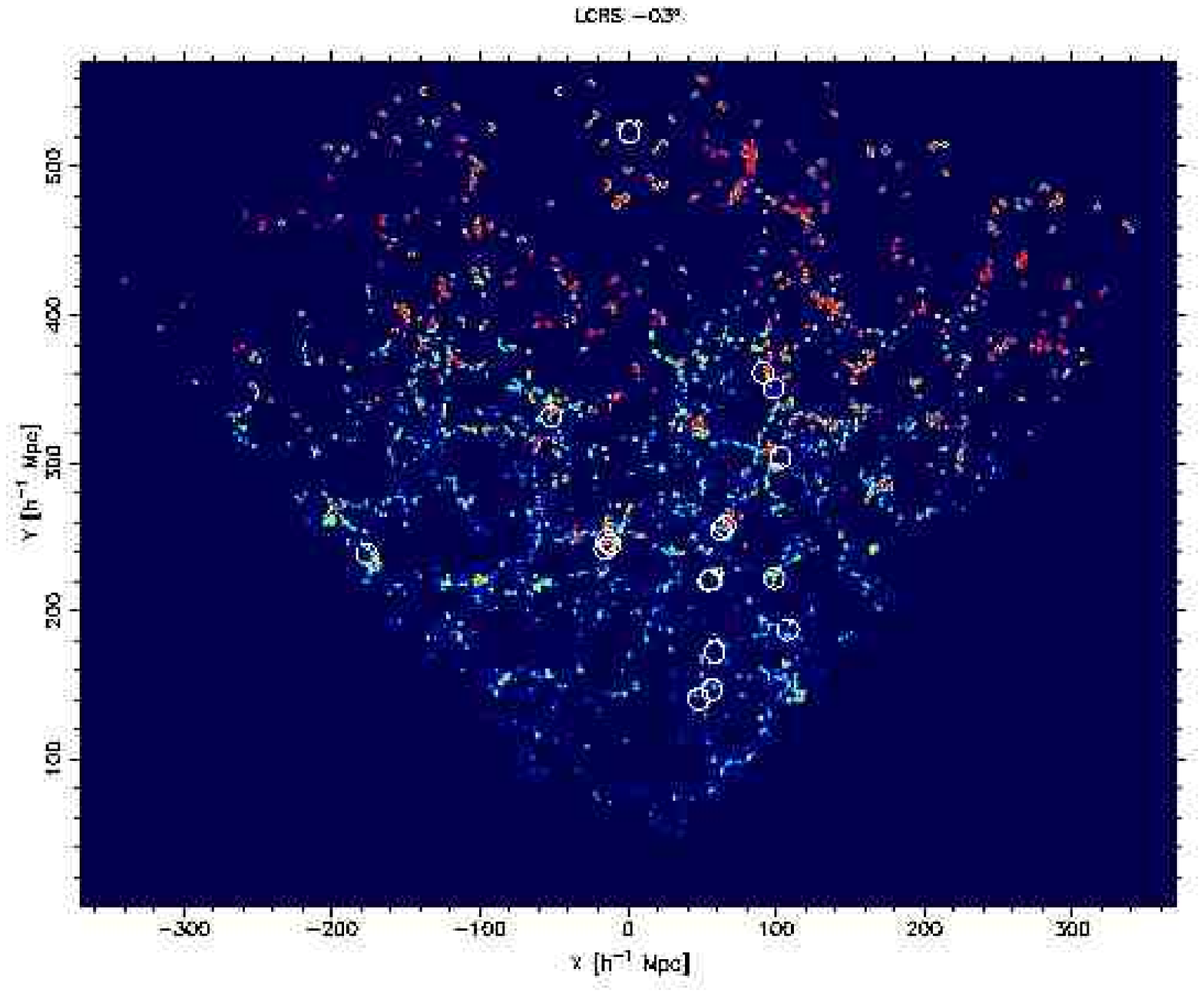}}\hspace{2mm}
\resizebox{0.45\textwidth}{!}{\includegraphics*{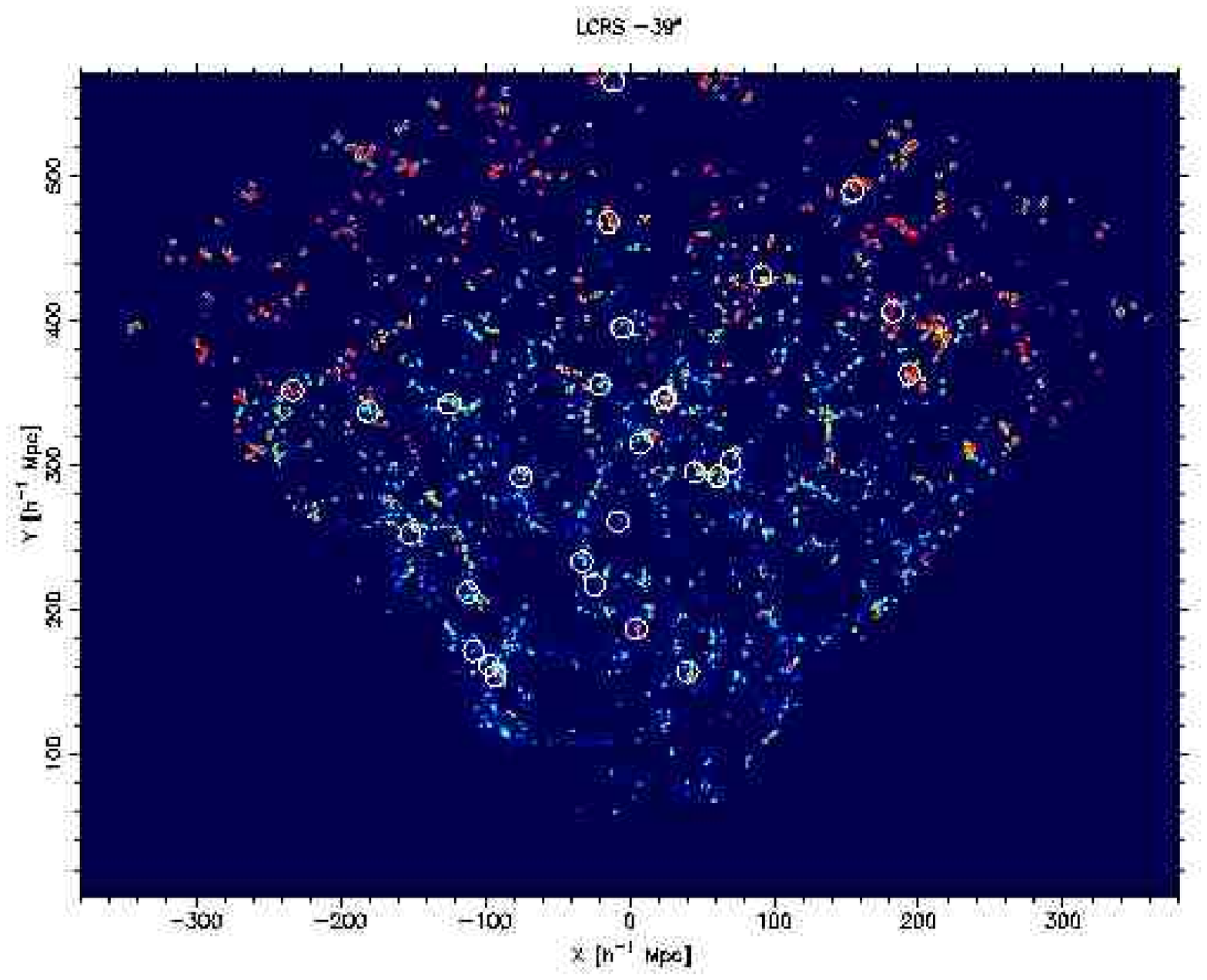}}\hspace{2mm}\\
\resizebox{0.45\textwidth}{!}{\includegraphics*{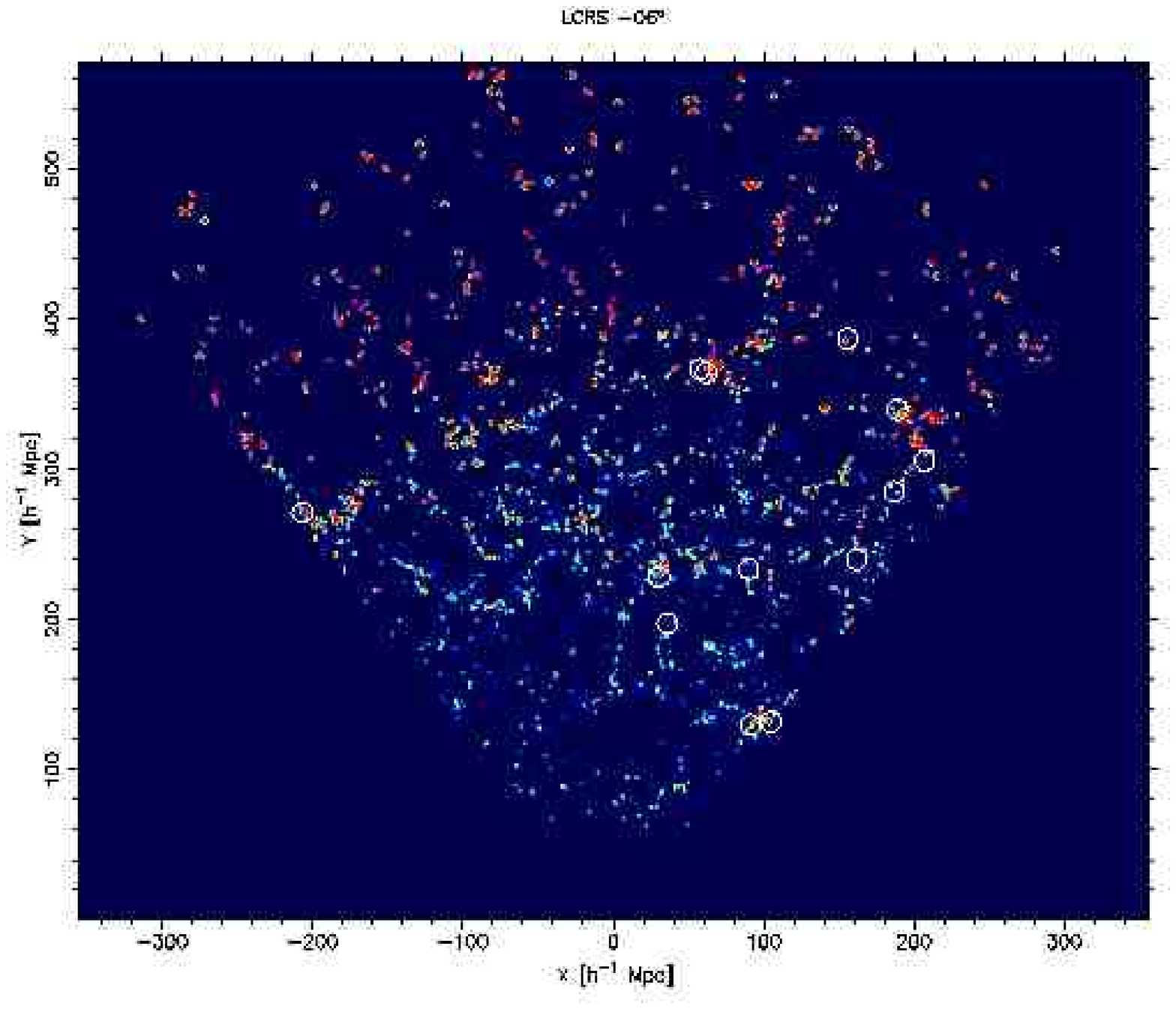}}\hspace{2mm}
\resizebox{0.45\textwidth}{!}{\includegraphics*{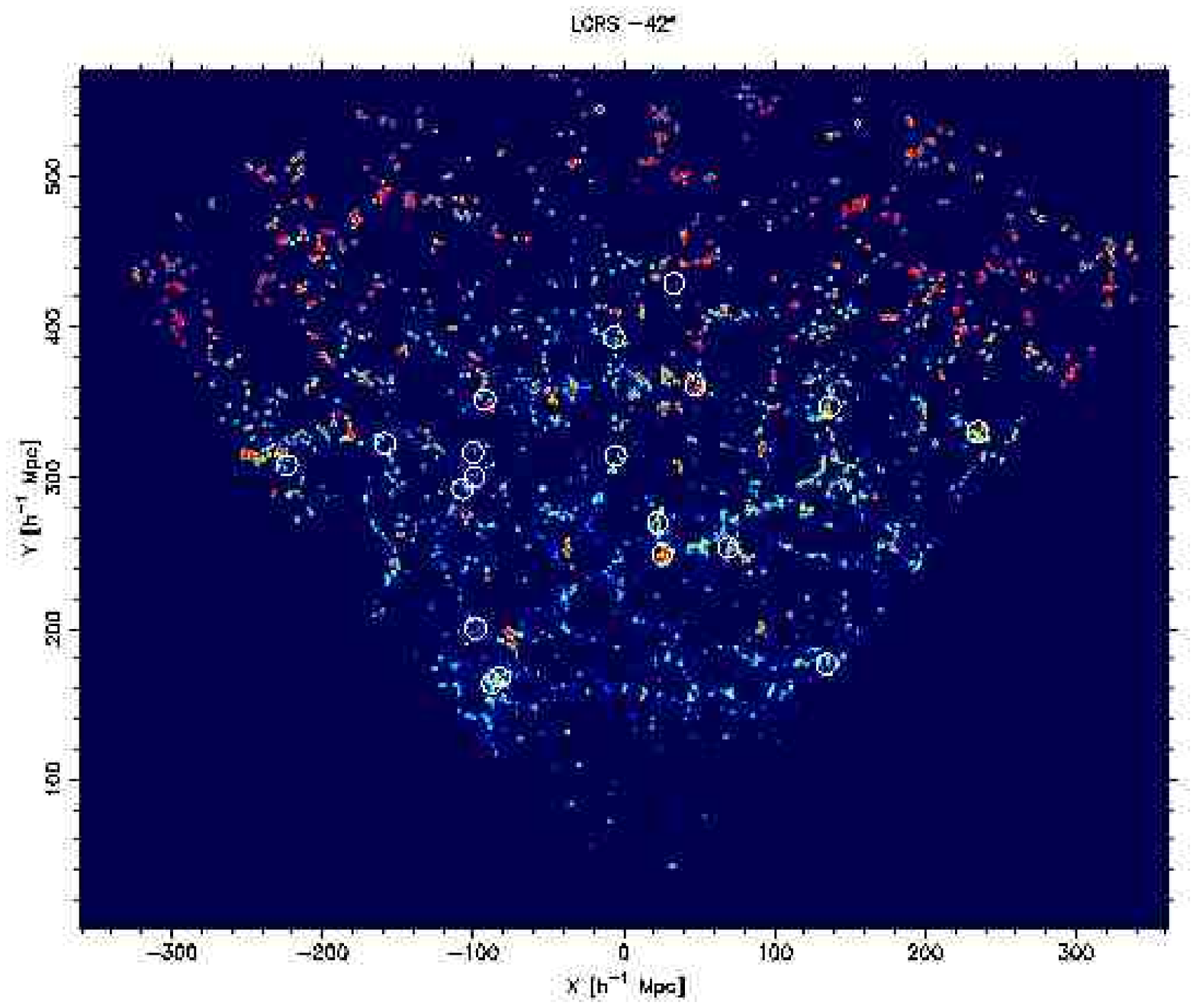}}\hspace{2mm}\\
\resizebox{0.45\textwidth}{!}{\includegraphics*{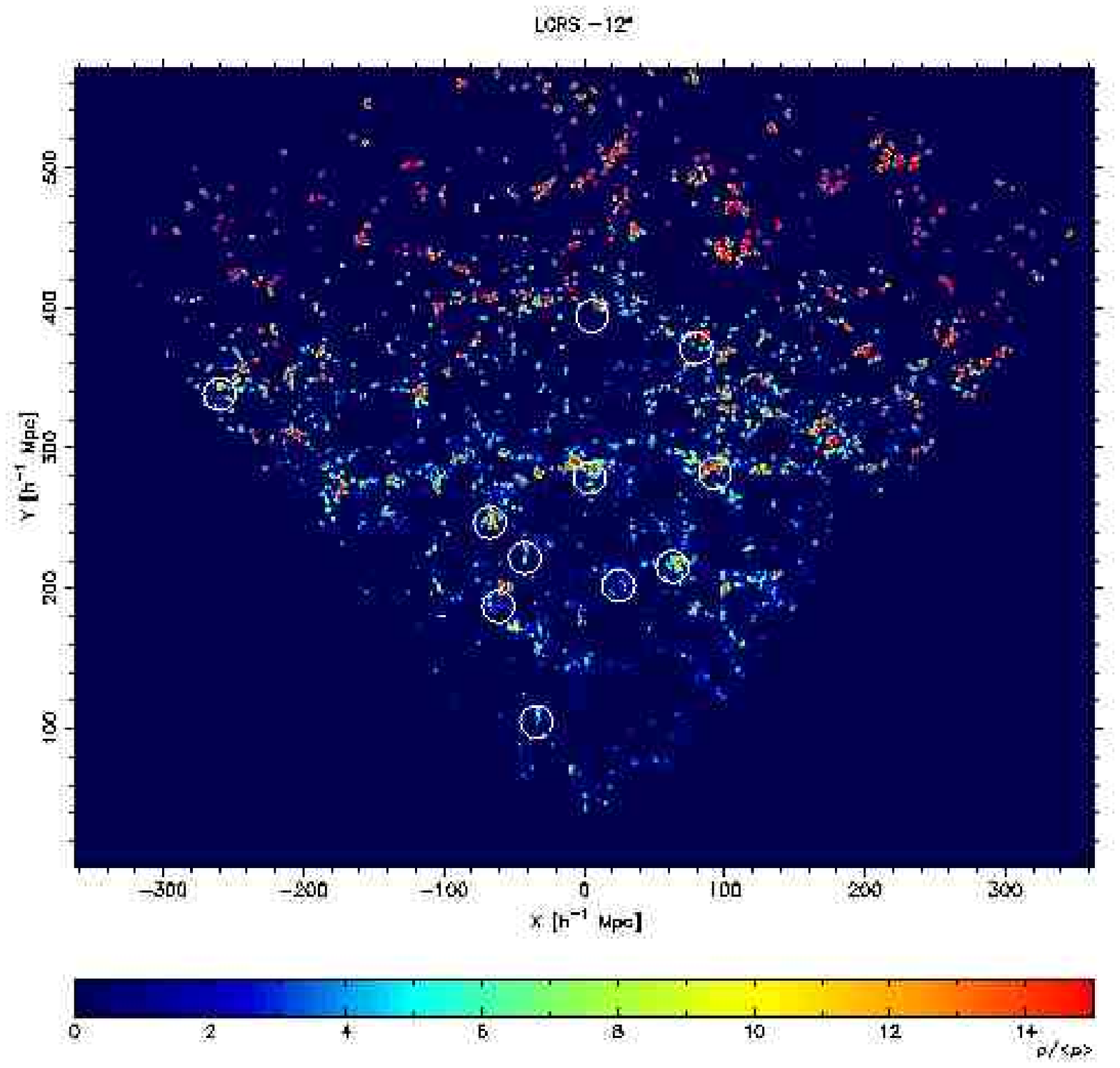}}\hspace{2mm}
\resizebox{0.45\textwidth}{!}{\includegraphics*{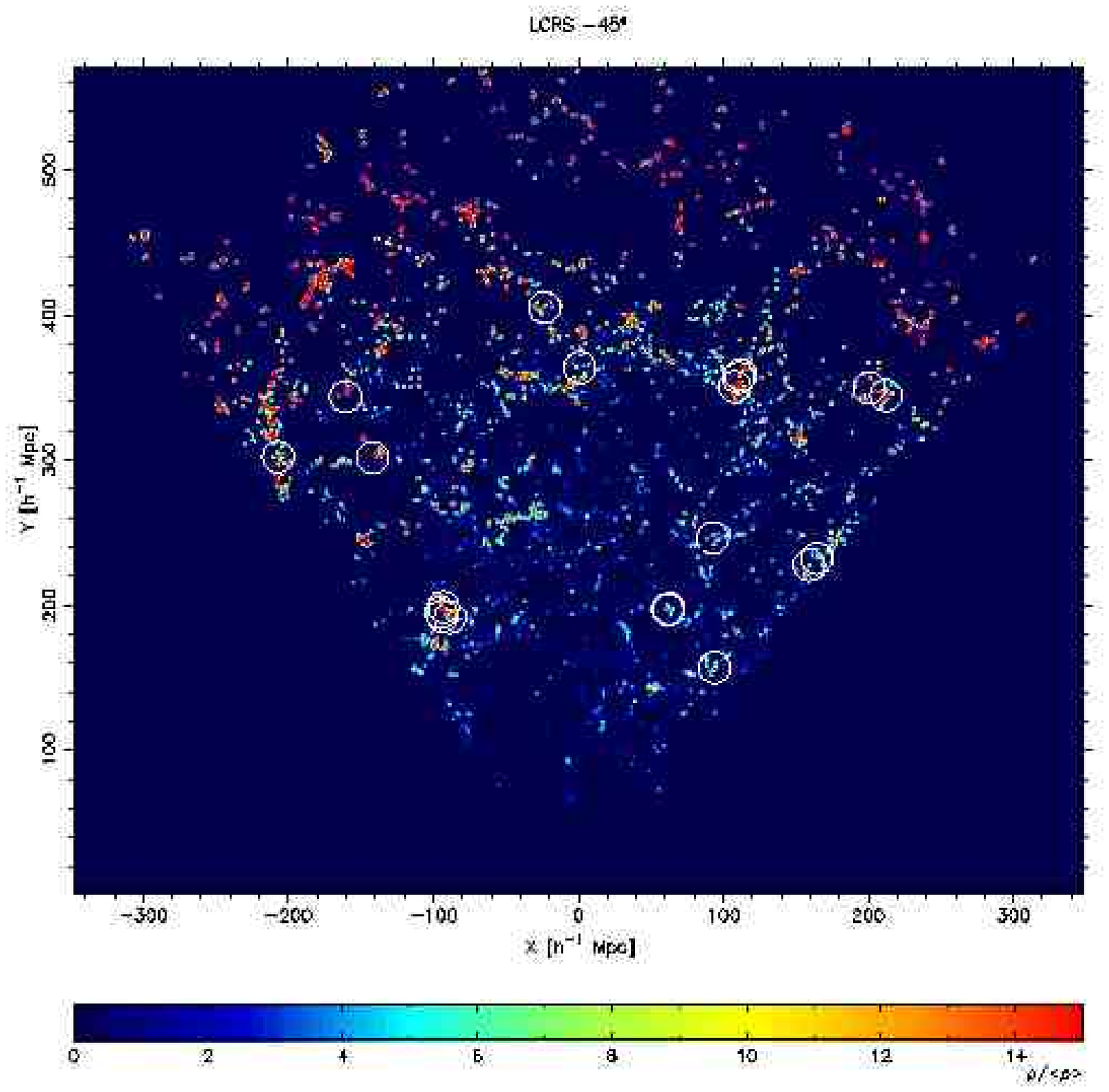}}\hspace{2mm}\\
\caption{The luminosity density field of the LCRS slices smoothed
  with a $\sigma = 0.8$~\Mpc\ Gaussian filter. Open circles denote
  positions of Abell clusters located within boundaries of slices. In
  some cases an Abell cluster consists of several subclusters, in
  these cases only rich subclusters are marked. The observer is
  located at the coordinates $(x,y)=(0,0)$.}
\label{fig:2}
\end{figure*}

\begin{figure*}[ht!]
\centering
\resizebox{0.45\textwidth}{!}{\includegraphics*{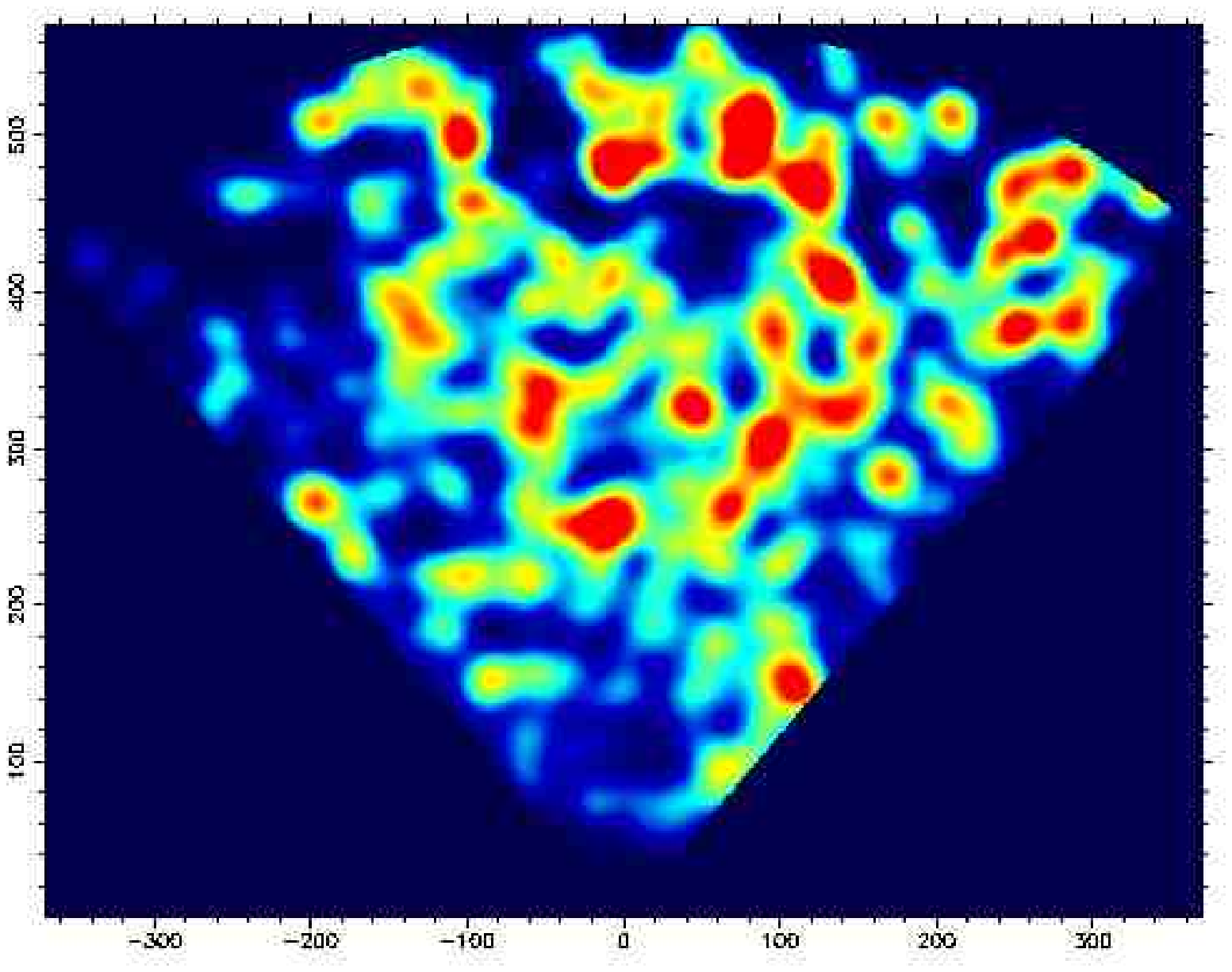}}\hspace{2mm}
\resizebox{0.45\textwidth}{!}{\includegraphics*{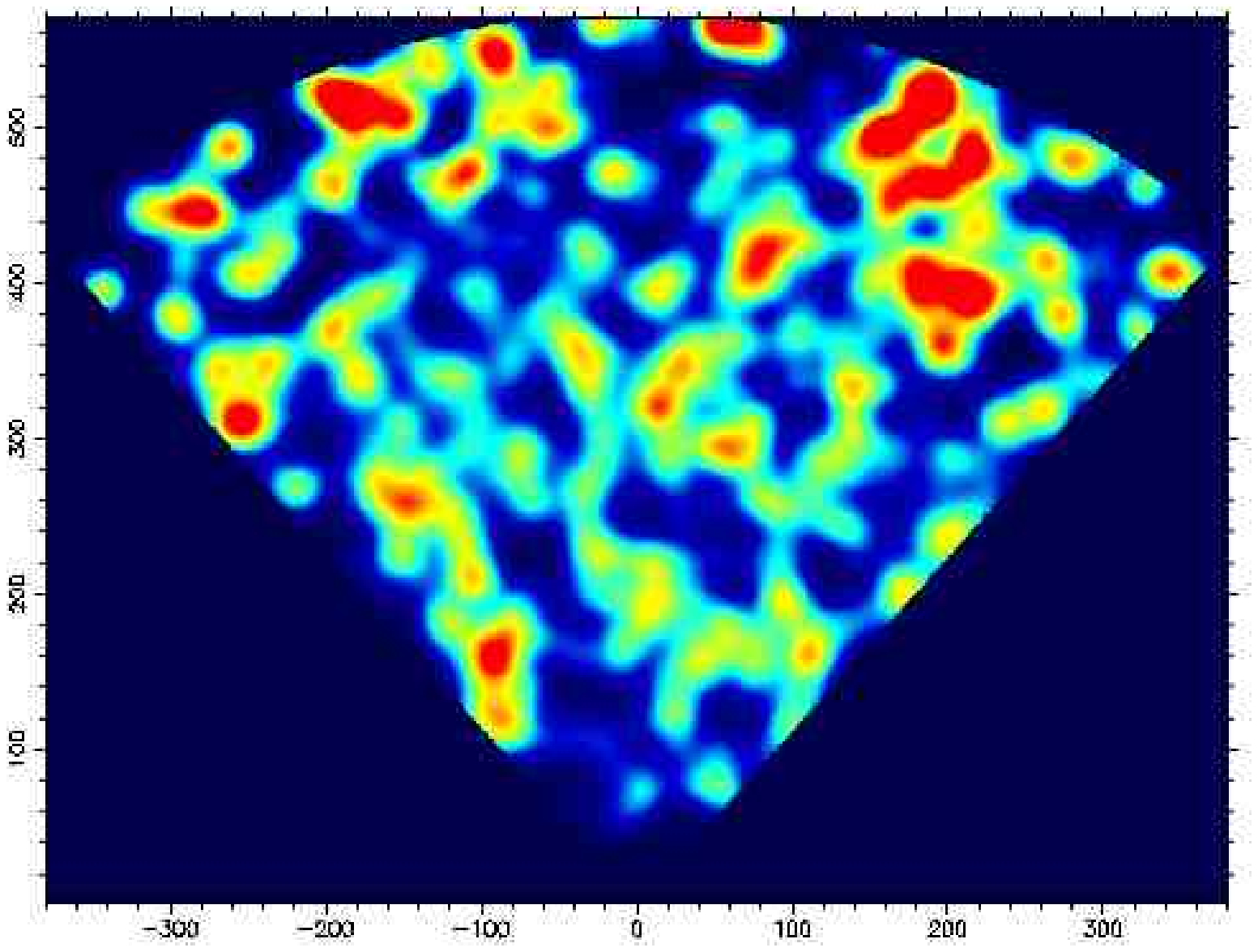}}\hspace{2mm}\\
\resizebox{0.45\textwidth}{!}{\includegraphics*{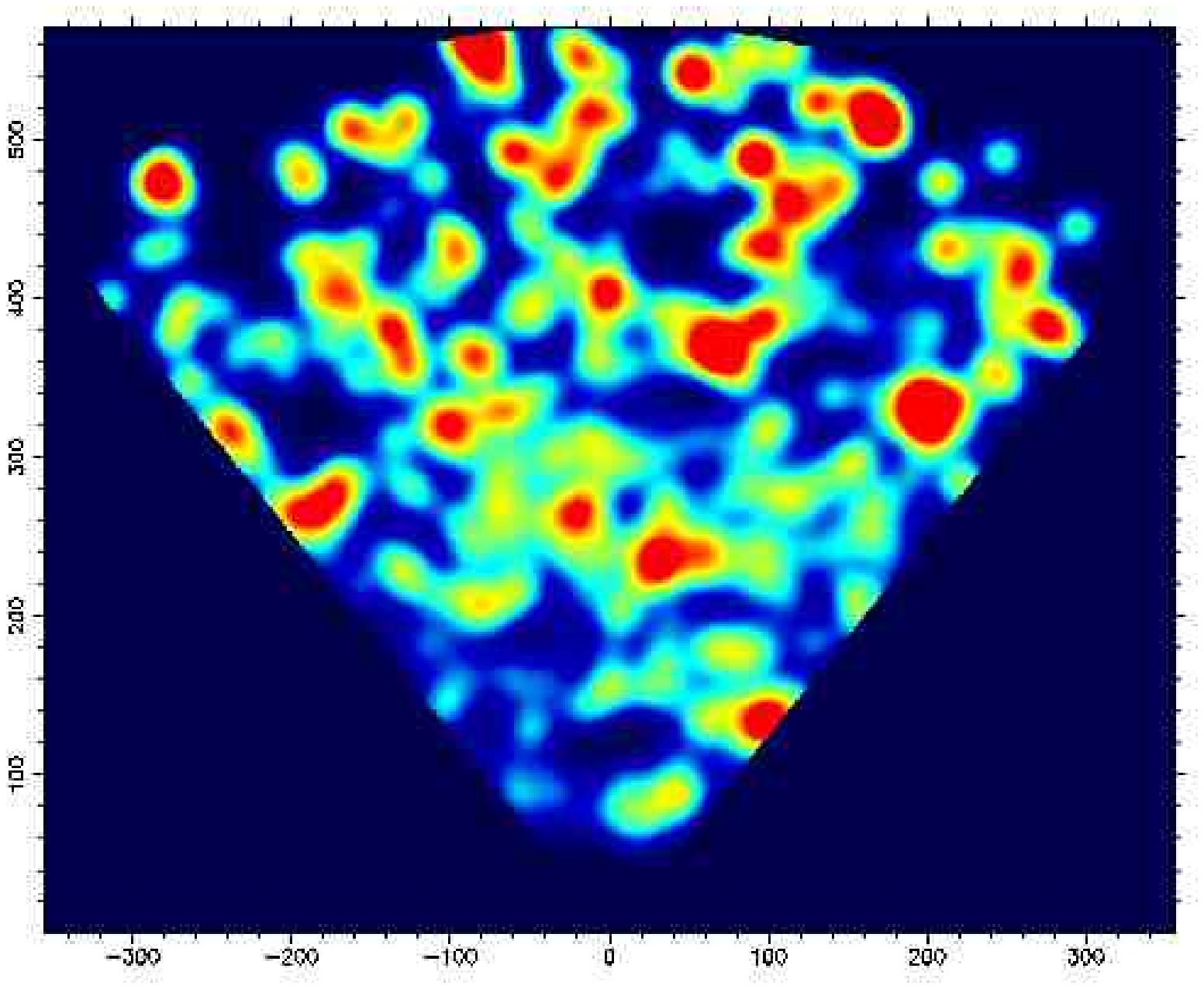}}\hspace{2mm}
\resizebox{0.45\textwidth}{!}{\includegraphics*{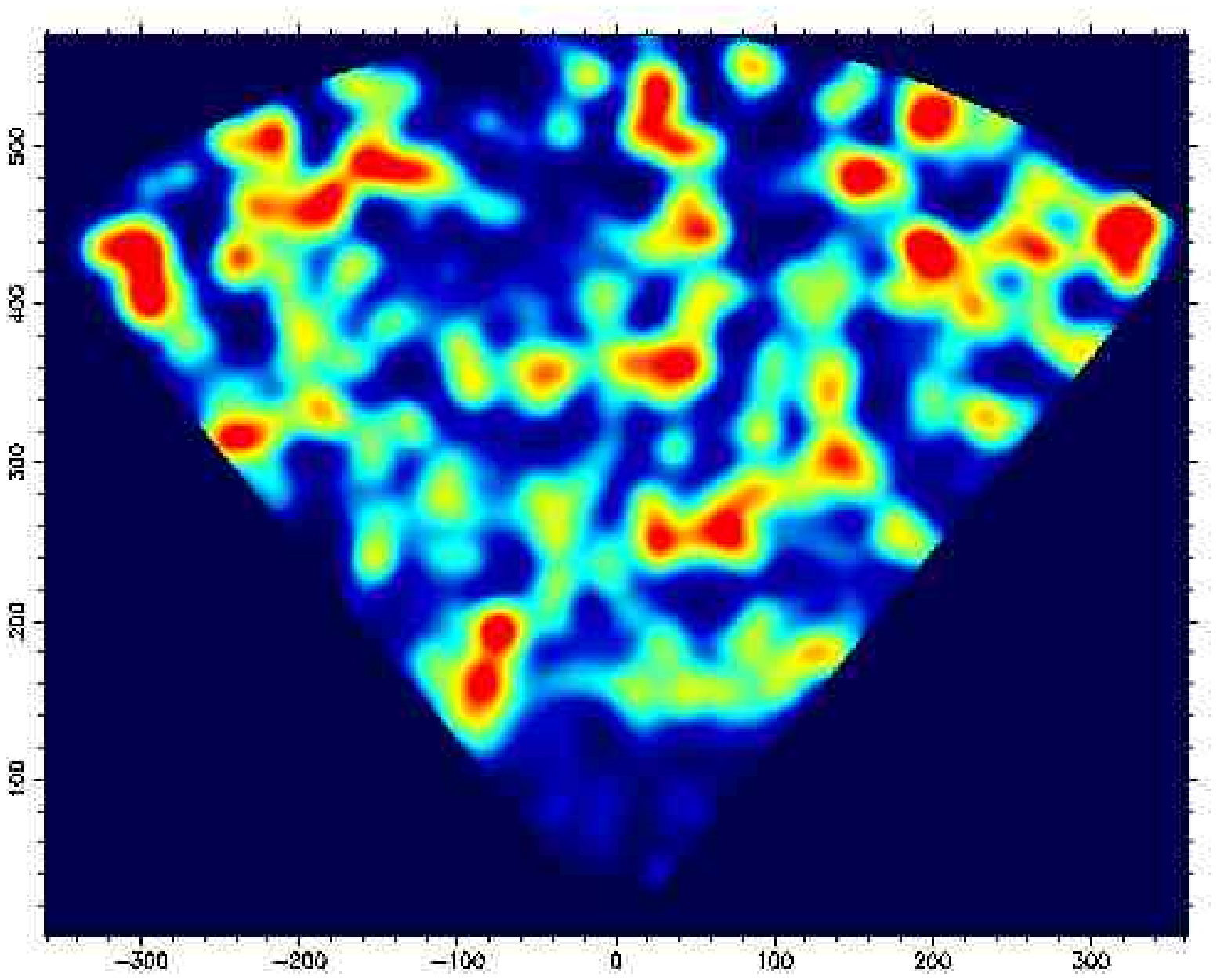}}\hspace{2mm}\\
\resizebox{0.45\textwidth}{!}{\includegraphics*{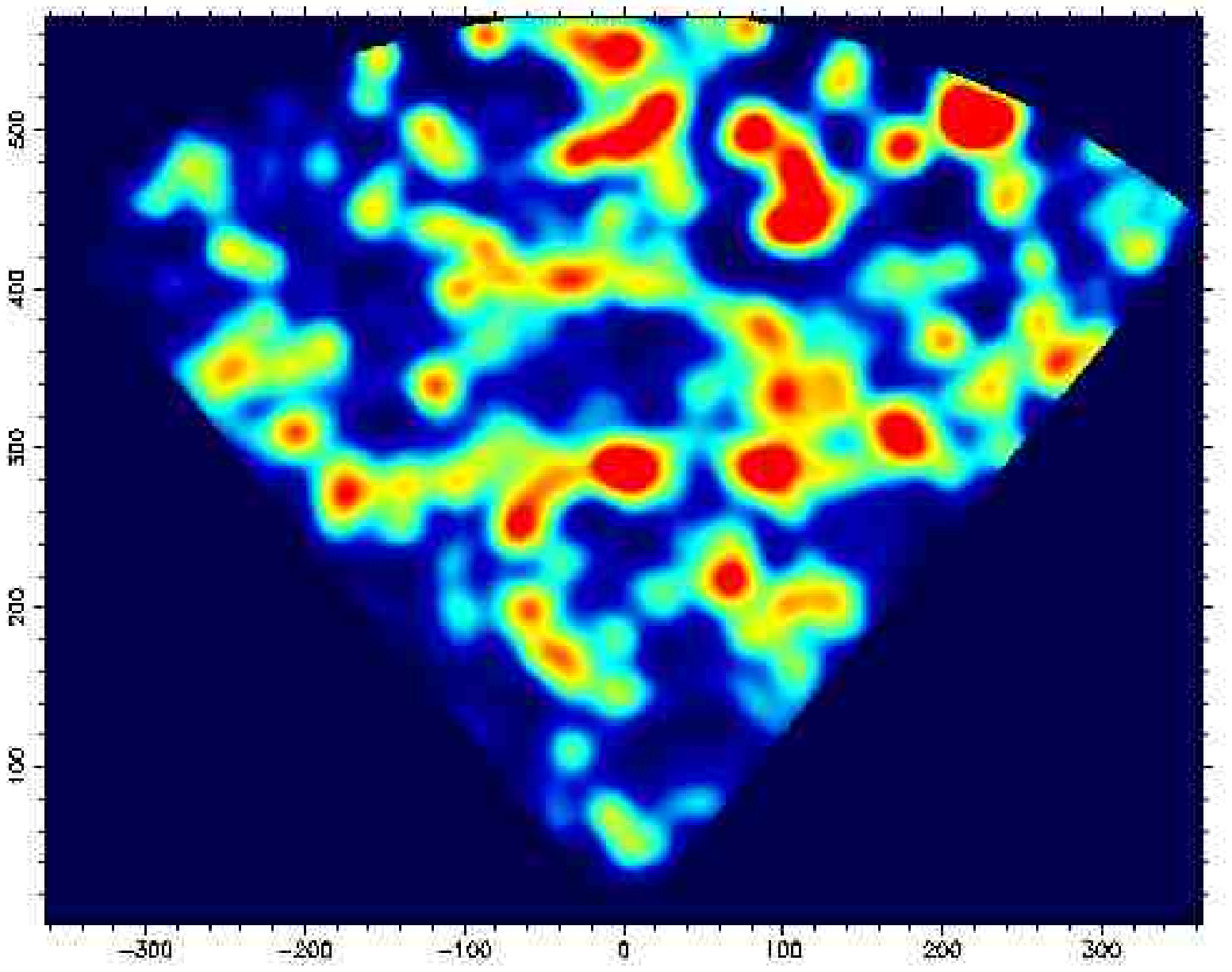}}\hspace{2mm}
\resizebox{0.45\textwidth}{!}{\includegraphics*{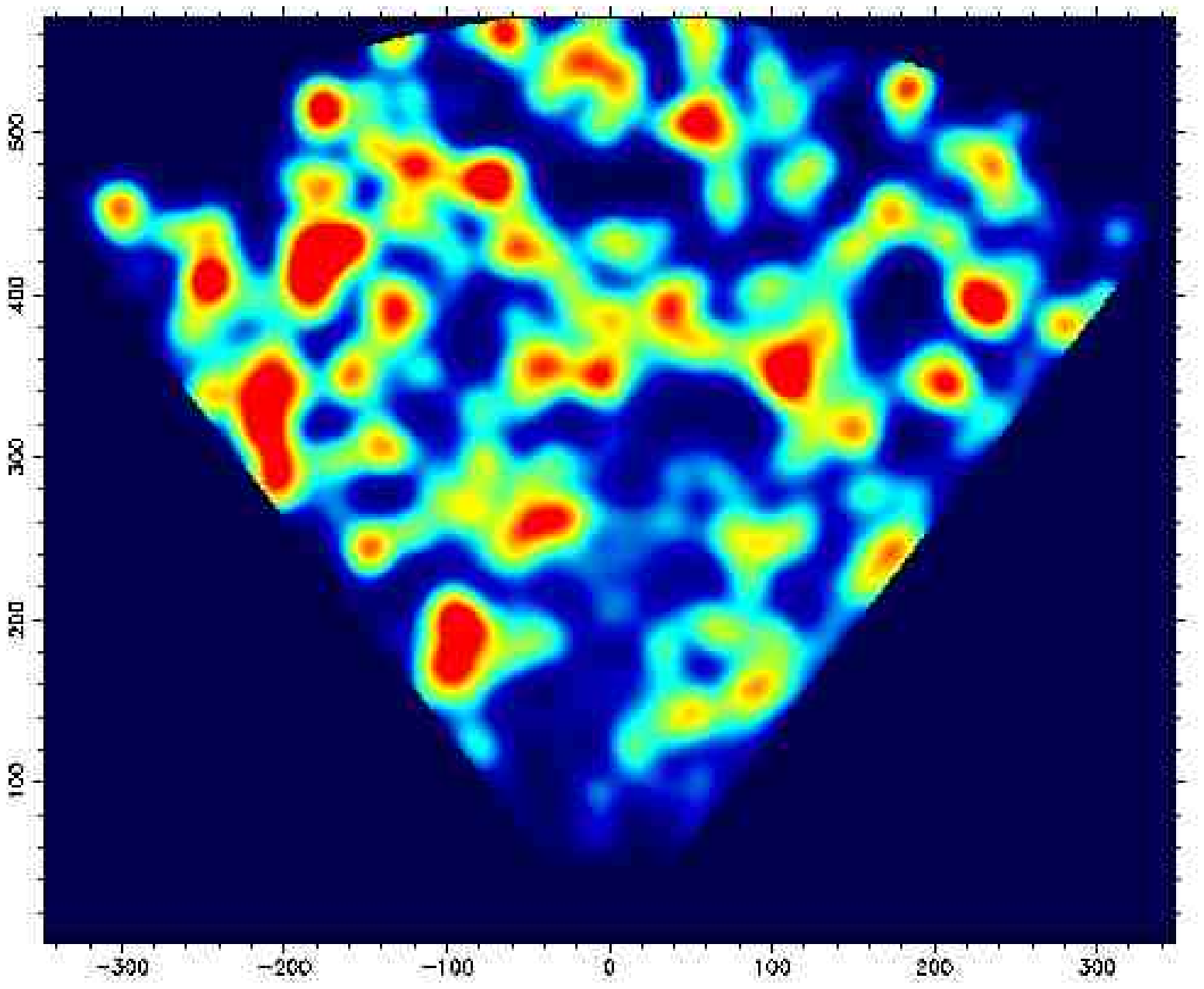}}\hspace{2mm}\\
\caption{The density field of the LCRS slices smoothed with a
$\sigma = 10$~\Mpc\ Gaussian filter. Panels are located as in Fig.~2.} 
\label{fig:3}
\end{figure*}

\section{Density field clusters}

\subsection{The DF-cluster catalogue}

We use the high-resolution density field to find compact overdensity
regions.  We call these regions density field clusters (DF-clusters).
The density field and the DF-clusters were found as follows.

First, we calculated the comoving distance for every LCRS galaxy using
a cosmological model with the matter density $\Omega_{\rm m} = 0.3$,
and the dark energy density (cosmological constant) of
$\Omega_{\Lambda} = 0.7$ (both in units of the critical cosmological
density).  In calculating absolute magnitudes we used the K-correction
and the correction for absorption in the Milky Way (for details see
H03).  To calculate the density field we used weights, which take into
account the expected luminosity of galaxies outside the visibility
window $m_1 \dots m_2$, using a procedure described in Paper I (see
also TUC). In doing so we assume that every galaxy is a visible member
of a density enhancement.  This density enhancement is actually a
halo, consisting of one or more bright galaxies in the visibility
window, and galaxies fainter or brighter than seen in the visibility
window. In calculating the total luminosity of the DF-cluster we
assume that luminosities of galaxies are distributed according to the
Schechter ~(\cite{Schechter76}) luminosity function. The estimated
total luminosity per a visible galaxy is
\begin{equation}
L_{tot} = L_{obs} W_L, 
\label{eq:ldens}
\end{equation}
where $L_{obs}=L_{\odot }10^{0.4\times (M_{\odot }-M)}$ is the
luminosity of the visible galaxy of absolute magnitude $M$, and 
\begin{equation}
W_L =  {\frac{\int_0^\infty L \phi
(L)dL}{\int_{L_1}^{L_2} L \phi (L)dL}} 
\label{eq:weight}
\end{equation}
is the weight -- the ratio of the expected total luminosity to the
expected luminosity in the visibility window. In the last equation
$L_i=L_{\odot }10^{0.4\times (M_{\odot }-M_i)}$ are the luminosities
of the observational window limits corresponding to the absolute
magnitudes $M_i$, and $M_{\odot }$ is the absolute magnitude of the
Sun.  In calculating the weights we used the values of the parameters
of the Schechter function, $\alpha$ and $M^{\ast}$, as found in H03
(and reproduced in Table~\ref{Tab2}).  Here N50, S50, N112, and
S112 denote the 50 and 112 fibre fields in the Northern and Southern
hemisphere, and NS112 is the estimate for all the 112 fibre fields.  In
calculating the weights we integrated instead of 0 to $\infty$ over an
absolute magnitude range from $M_0 = -13.0$ to $M_{lim} = -24.5$
in the R-photometric system.

\begin{figure*}[ht]
\vspace*{8.0cm}
\caption{The luminosity density of the LCRS slices as a function of
distance. The left panel shows Northern slices, the right panel shows
Southern slices. } 
\includegraphics{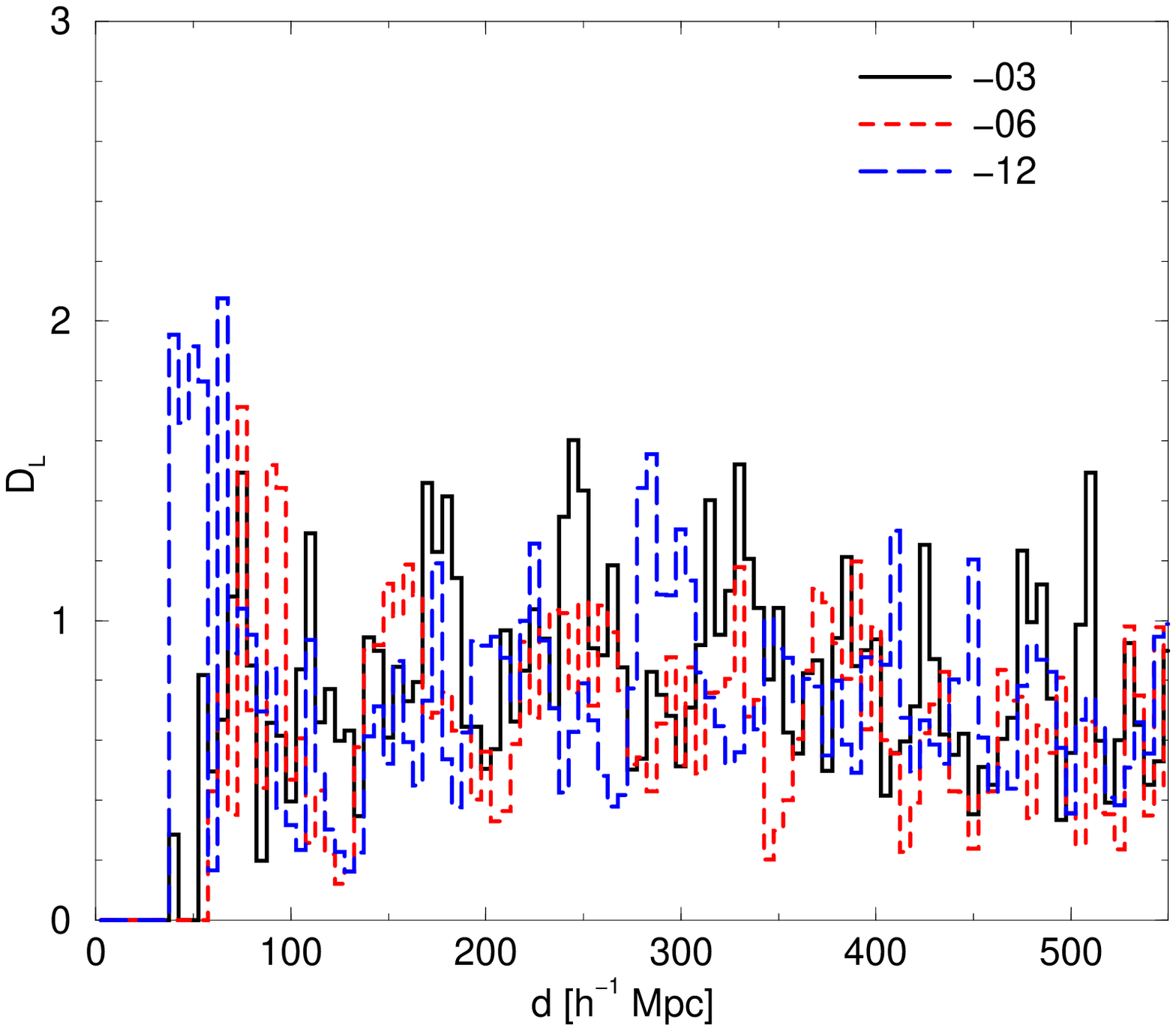}
\includegraphics{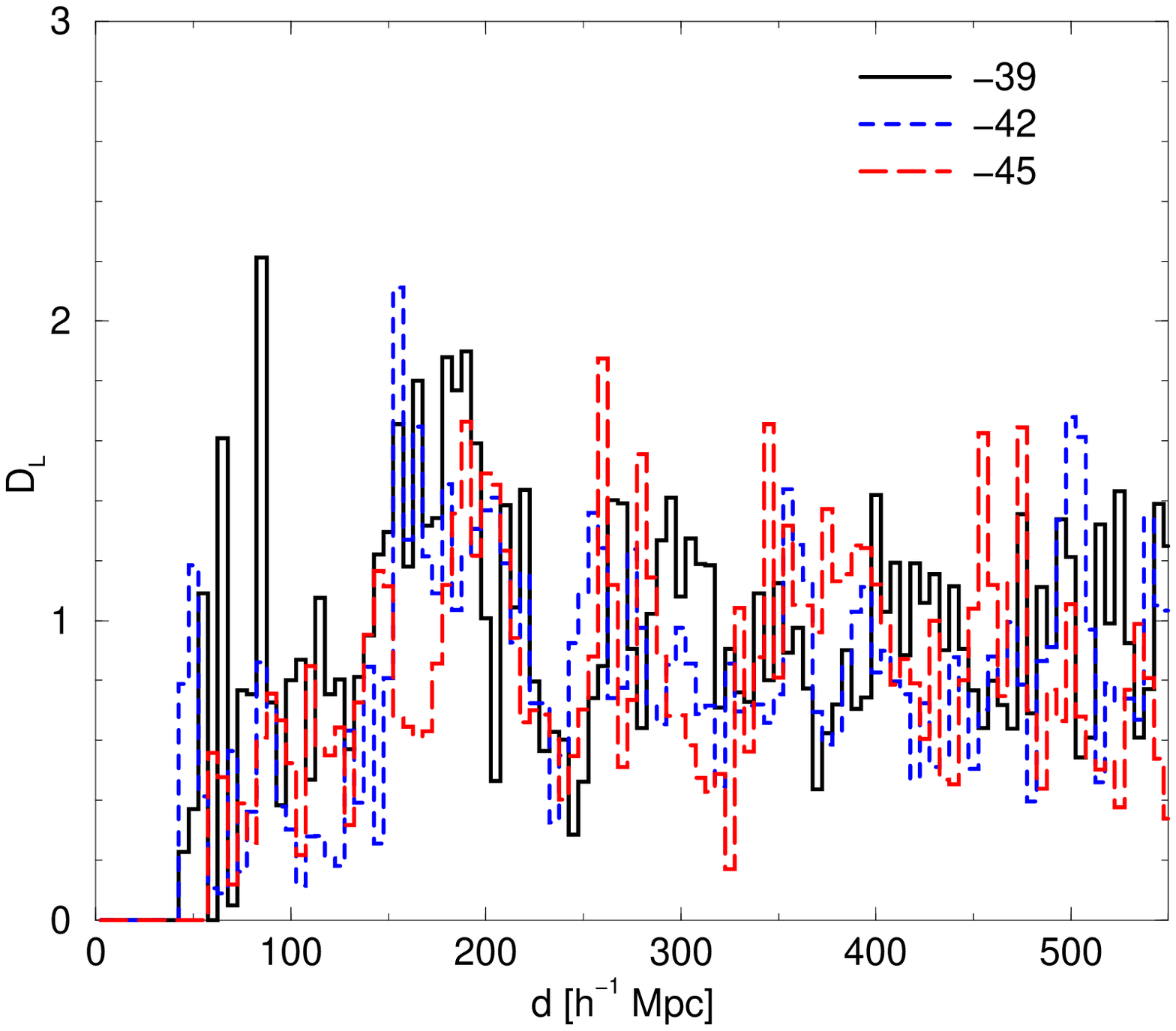}
\label{fig4}
\end{figure*}

\begin{table}[ht]
\caption{The best fitting $M^{*}$ and $\alpha$ for the LCRS samples}
\label{Tab2}
\begin{tabular}{ccc}
\hline\hline
Sample & $M^{*}-5\log h$ & $\alpha $ \\ \hline
N50 & $-\left( 20.33\pm 0.12\right) $ & $-\left( 0.40\pm 0.18\right) $ \\ 
S50 & $-\left( 20.64\pm 0.18\right) $ & $-\left( 0.74\pm 0.21\right) $ \\ 
N112 & $-\left( 20.40\pm 0.05\right) $ & $-\left( 0.76\pm 0.07\right) $ \\ 
S112 & $-\left( 20.40\pm 0.05\right) $ & $-\left( 0.70\pm 0.07\right) $ \\ 
NS112 & $-\left( 20.38\pm 0.04\right) $ & $-\left( 0.70\pm 0.04\right) $ \\ 
TOTAL & $-\left( 20.40\pm 0.03\right) $ & $-\left( 0.69\pm 0.04\right) $ \\ 
\hline
\end{tabular}
\end{table}

We plot in Fig.~\ref{fig:1} the absolute magnitudes of the window,
$M_1$ and $M_2$, corresponding to the observational window of apparent
magnitudes at the distance of the galaxy, and observed absolute
magnitudes of galaxies, $M_{obs}$.  We also plot in Fig.~\ref{fig:1} 
the estimated total luminosity per visible galaxy (in units of $10^{10}$
solar luminosities) for the $-3^{\circ}$ slice galaxies as a function of
distance.  This total luminosity was used in calculating the density
field.

Fig.~\ref{fig:1} shows that the observational window limits $M_1$ and
$M_2$ form several strips in the magnitude--distance diagram. This is
due to differences in the apparent magnitude window of the 50 and
112-fibre fields (in particular, in the bright end of the window,
where we have several parallel strips of the limit in the left panel
of Fig.~\ref{fig:1}), as well as other observational selection effects
discussed by TUC (which increase the width of strips).  These
differences have been taken into account in the calculation of the
luminosity function to find total luminosities for galaxies, and as a
result we see no strips in the distribution of total luminosities,
plotted in the right panel of Fig.~\ref{fig:1}.

Next we smoothed the density field with a Gaussian filter of a
smoothing length 0.8~\Mpc.  As described in Paper I, in calculating 
the density field we used a 2-dimensional grid with a cell size
1~\Mpc.  This yields a high-resolution map where the individual density
enhancements can be easily recognised. This high-resolution density
field is presented in Fig.~\ref{fig:2}.  Fig.~\ref{fig:3} presents the
low-resolution density field found using a 10~\Mpc\ smoothing
length. We used this field to find DF-superclusters and to define the
global density, characterising the environment of DF-clusters (see
Sect. 3.3 below).  The high-resolution maps show the density
distribution in wedges of increasing thickness as the distance from
the observer increases.  The low-resolution density maps are converted
to sheets of constant thickness by dividing the surface density to the
thickness of the sheet at particular distance from the observer.

To identify DF-clusters, every cell of the field was examined to see
whether its density exceeds the density of all neighbouring cells.  If
the density of the cell was higher than that of all its neighbours,
then the cell was considered to be the centre of a DF-cluster. The total
luminosity of the DF-cluster was determined by summing luminosity
densities of cells within a box of size $-2 \leq \Delta x \leq 2$, and
$-2 \leq \Delta y \leq 2$ in cell size units. This range corresponds
to the smoothing length 0.8~\Mpc\ which distributes the luminosity of
every galaxy between the central and 24 neighbouring cells.  The
luminosities were calculated in solar luminosity units.  At large
distances the LCRS sample is rather diluted, and there are only a few
galaxies in the nearby region of the LCRS slices. Thus we included
into our catalogue of DF-clusters only objects within the distance
interval $100 \dots 450$~\Mpc.  The DF-cluster sample has only a few
low-luminosity clusters; thus we included in our catalogue only
clusters having total luminosities over $L_0 \geq 0.5\times 10^{10}
L_{\odot}$.  The number of DF-clusters found in the individual slices
is given in Table~1.

According to the general cosmological principle the mean density of
luminous matter (smoothed over superclusters and voids) should be the
same everywhere.  A weak dependence on distance may be due to
evolutionary effects: luminosities of non-interacting galaxies
decrease as stars age. If we ignore this effect we may expect that the
total corrected luminosity density should not depend on the distance
from the observer, in contrast to the number of galaxies which is
strongly affected by selection (for large distances we do not see
absolutely faint galaxies). This difference in observed and total
luminosity is clearly seen in Fig.~\ref{fig:1}: with increasing distance
total luminosities exceed observed ones by a factor of ten or more. We
can use the mean luminosity density as a test of our weighting
procedure.  In Fig.~\ref{fig4} we show the mean luminosity density in
spherical shells of thickness 5~\Mpc\ for all 6 slices of the LCRS.
We see strong fluctuations of the luminosity density, caused by
superclusters and voids.  The overall mean density is, however, almost
independent of the distance from the observer.  The mean density is a
very sensitive test for the parameters of the luminosity function. It
shows that the presently accepted set of parameters of the luminosity
function compensates correctly the absence of faint galaxies in our
sample.

\begin{figure*}[ht]
\centering
\resizebox{0.45\textwidth}{!}{\includegraphics*{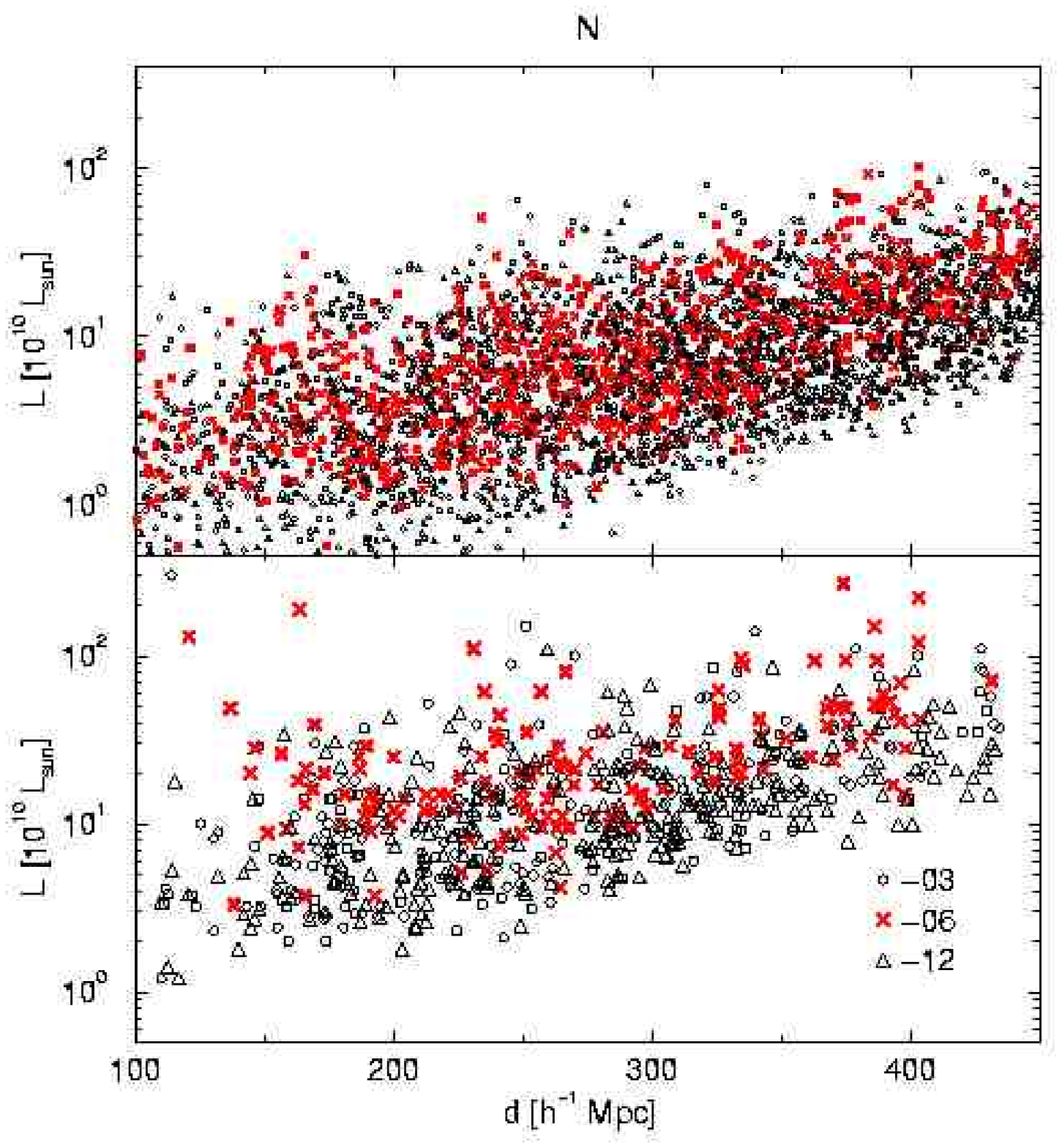}}\hspace{2mm}
\resizebox{0.45\textwidth}{!}{\includegraphics*{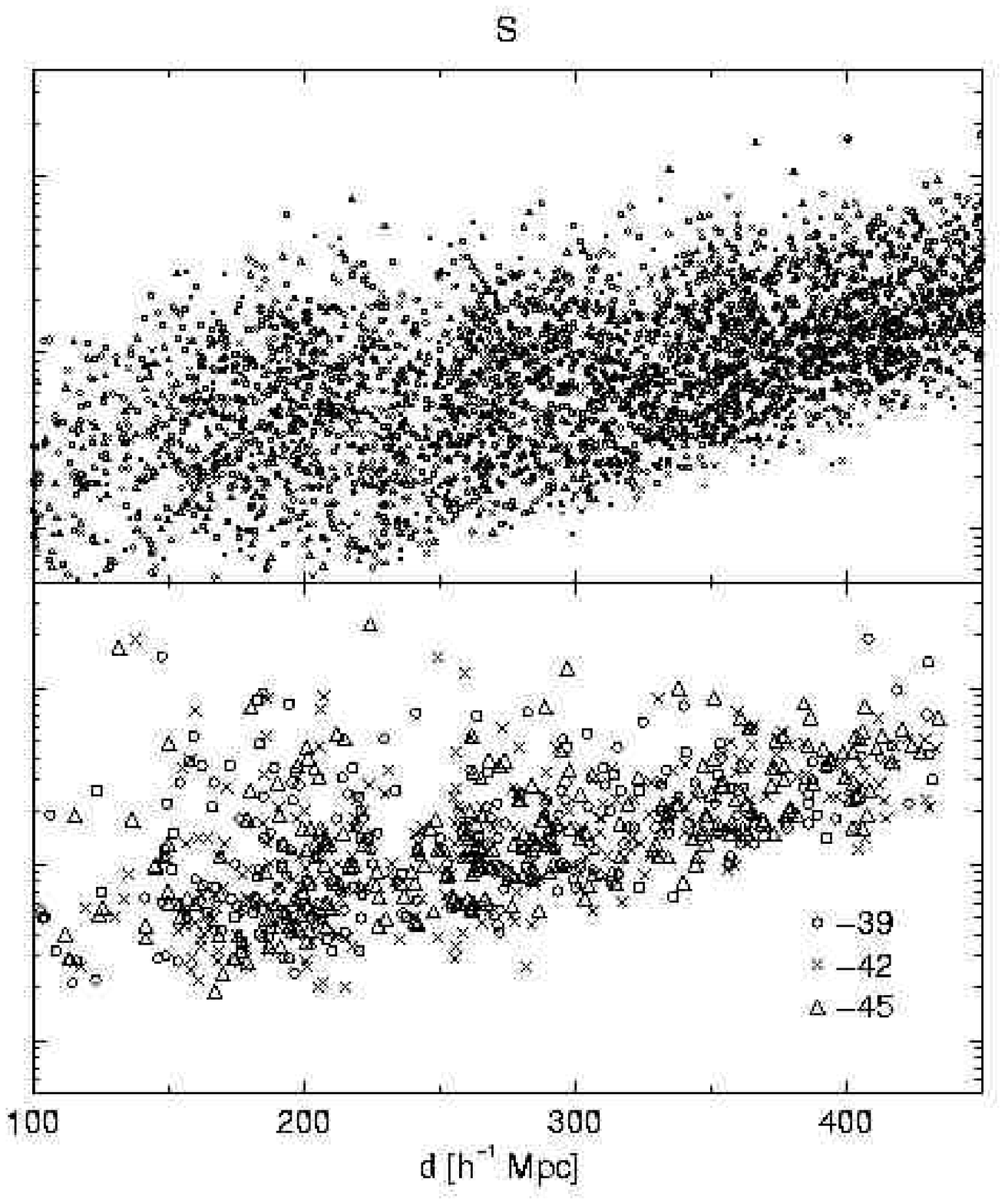}}\hspace{2mm}
\caption{The luminosities of DF-clusters as a function of
  distance. The upper panels show the distribution for the
  DF-clusters, the lower panels for the LCRS loose groups; the left
  panels show Northern slices, the right panels show Southern
  slices. }
\label{fig:5}
\end{figure*}

\begin{figure*}[ht]
\centering
\resizebox{0.45\textwidth}{!}{\includegraphics*{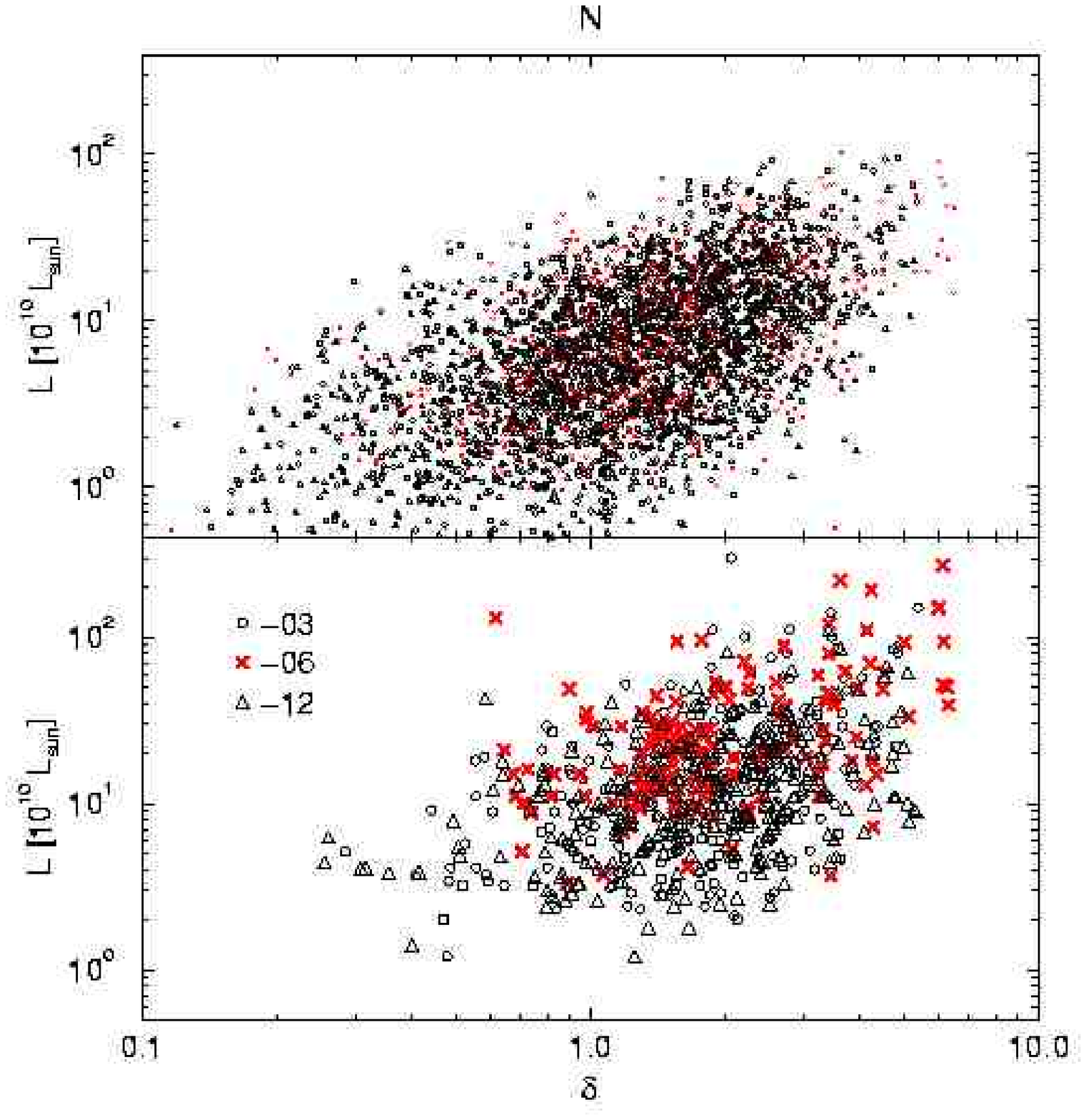}}\hspace{2mm}
\resizebox{0.45\textwidth}{!}{\includegraphics*{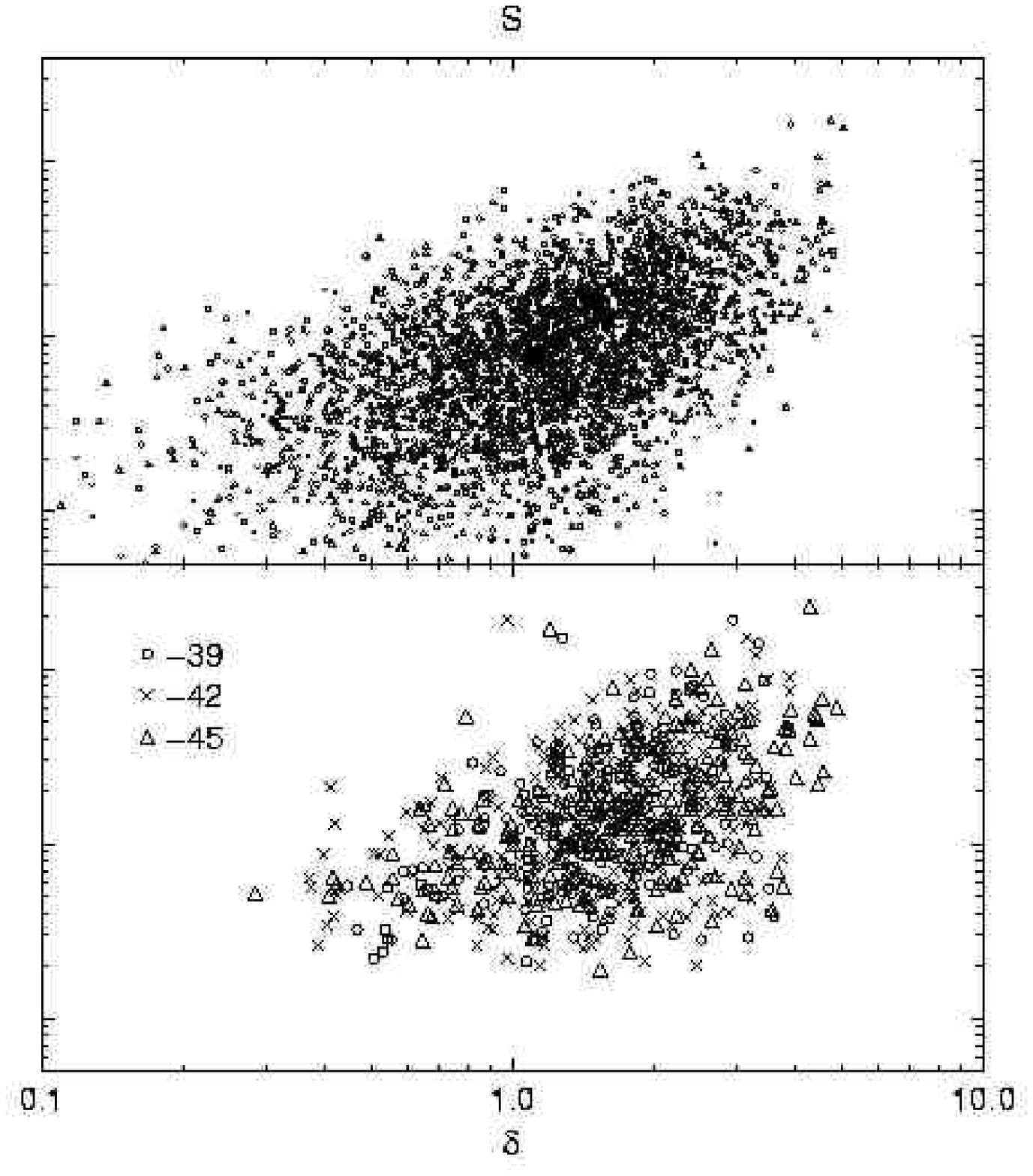}}\hspace{2mm}
\caption{Total luminosities of the DF-clusters (upper panels) and the
  LCRS loose groups (lower panels) as a function of the global relative
  density $\delta$. The left panels show Northern slices, the right
  panels show Southern slices. }
\label{fig:6}
\end{figure*}

\begin{figure*}[ht]
\centering
\resizebox{0.45\textwidth}{!}{\includegraphics*{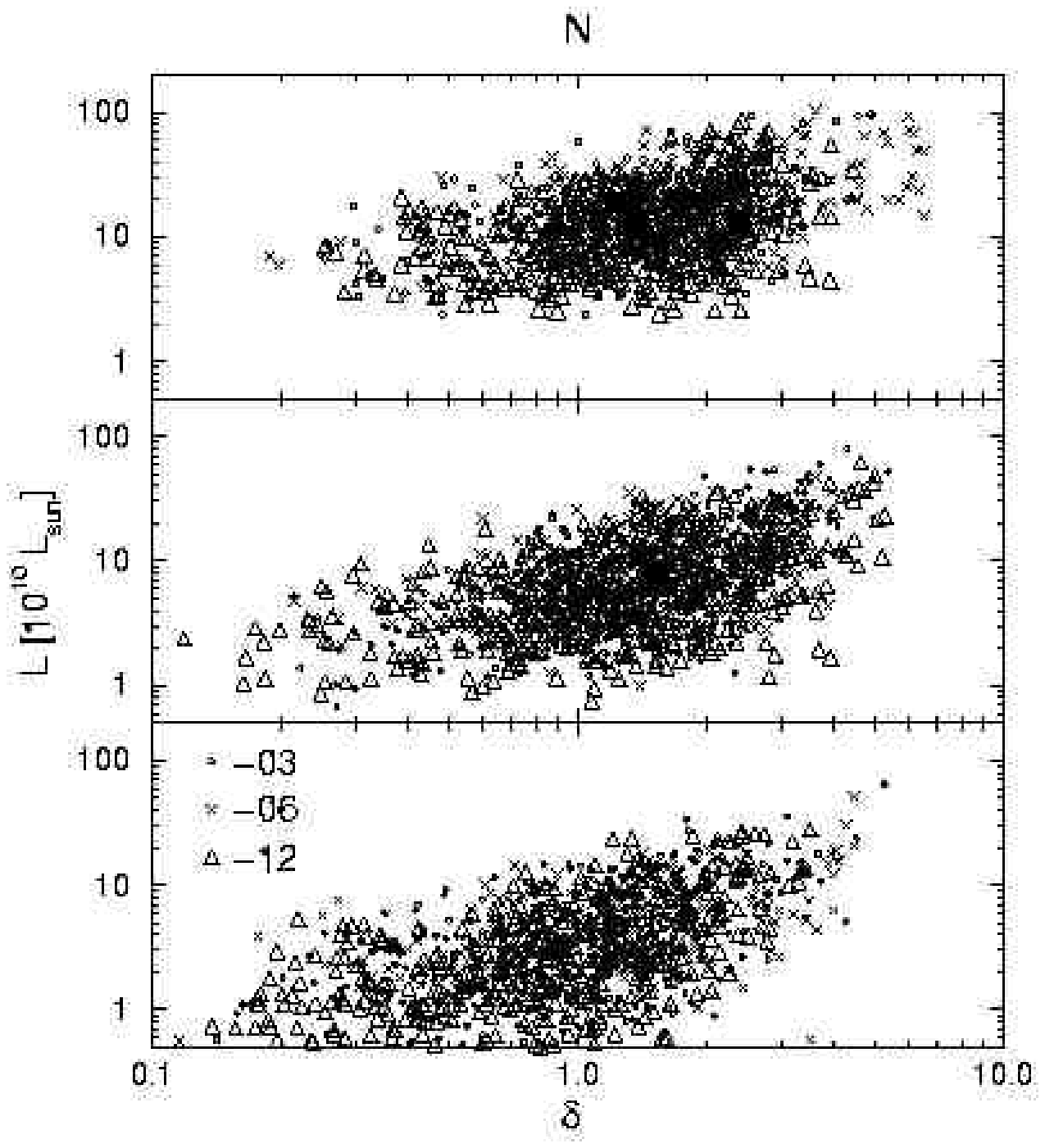}}\hspace{2mm}
\resizebox{0.45\textwidth}{!}{\includegraphics*{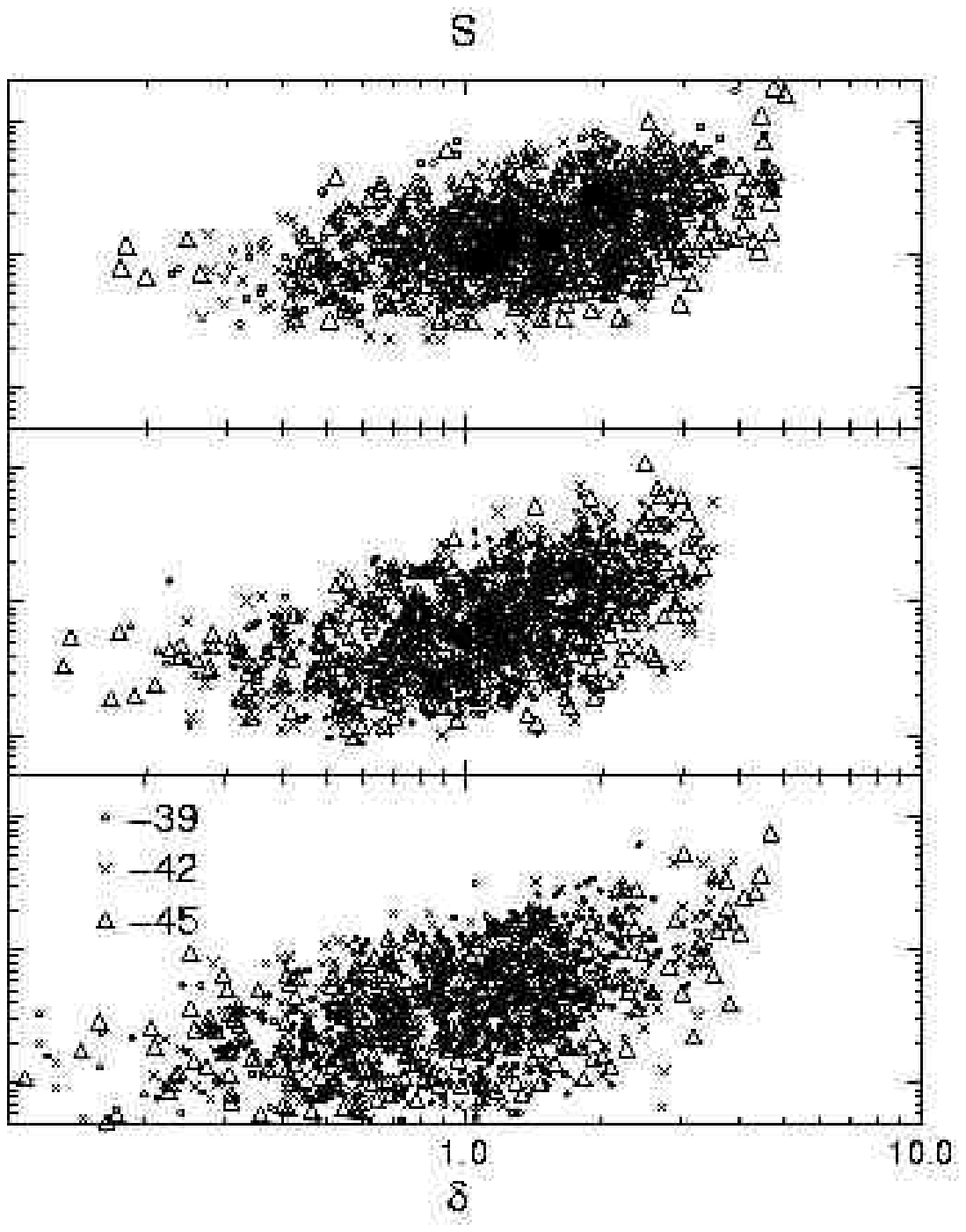}}\hspace{2mm}
\caption{Total luminosities of DF-clusters as a function of the global
relative density $\delta$; clusters are divided into 3 distance
classes: $100 \dots 250$, $250 \dots 350$, and $350 \dots 450$~\Mpc,
shown in the lower, middle and upper panels, respectively. The left
panels show Northern slices, the right panels show Southern slices. }
\label{fig:7}
\end{figure*}

\subsection{Selection effects}

The main selection effects in the LCRS (as in the SDSS) are due to the
finite width of the apparent magnitude window, $m_1 \dots m_2$, which
excludes galaxies outside this window from the
redshift survey.  This effect reduces the number of galaxies observed
for a given structure element (cluster) of the universe.  If the
cluster contains at least one galaxy within the visibility window of
the survey, then the contribution of the remaining galaxies to the
expected total luminosity of the cluster can be restored using the
weighting scheme discussed above.  However, if the cluster has no
galaxies in the visibility window, it is lost. For this reason, with
increasing distance from the observer, more and more mostly poor
clusters disappear from our survey.  This effect is clearly seen in
Fig.~\ref{fig:5}, which shows the total luminosities of DF-clusters as
a function of the distance from the observer, $d$.  For comparison we
also show the relationship between the luminosities and distances of
the LCRS loose groups of galaxies.  We see that low-luminosity clusters
are seen only at distances $d \le 250$~\Mpc. This limit is the same for
the DF-clusters and the LCRS loose groups, with the difference that
there are practically no LCRS loose groups with luminosities less than
$2\times 10^{10}~L_{\odot}$, whereas the lower limit of the
DF-clusters is $0.5\times 10^{10}~L_{\odot}$, i.e. 4 times lower.

There exists a well-defined lower limit of cluster luminosities at
larger distances; this limit is practically linear in the $\log L - d$
plot.  Within random fluctuations the lower luminosity limit is
identical for most LCRS slices: at 200 and 400~\Mpc\ it is 0.5 and
$4.8\times 10^{10}~L_{\odot}$, respectively; only the slice
$-6^{\circ}$ has a factor of 2 higher limit. This slice was observed
with 50 fibres only, and has a narrower apparent magnitude window. The
LCRS loose group sample has at 200 and 400~\Mpc\ a completeness
limit of 2 and $16\times 10^{10}~L_{\odot}$, respectively, i.e. a factor
of $\sim 3.3$ higher than that for the DF-cluster sample.  The absence
of low-luminosity clusters at large distances can be taken into
account statistically in the calculation of the cluster luminosity
function (see below).  The location of these missing clusters is not
known.  Thus with increasing distance there are fewer poor clusters to
trace the large-scale structure.

The more luminous DF-clusters and the LCRS loose groups form
volume-limited cluster samples; the number of clusters in these
samples is, however, considerably smaller than in the full samples.
Moreover, the exclusion of poorer clusters would make the
investigation of the dependence of cluster richness on environment
difficult. The study of the internal structure of superclusters and
voids would also be difficult. Thus we have not used volume-limited
subsamples of clusters.

In addition to the above selection effect the LCRS has one more
problem: due to relatively small number of fibres used in measuring
redshifts of galaxies the samples were diluted, i.e. not all galaxies
within the observational window $m_1 \dots m_2$ were observed for
redshifts.  This effect is strong in the $-6^{\circ}$ slice, which was
observed only with the 50-fibre spectrograph.  For this reason, the
number of loose groups detected by TUC in this slice is only about 
half that of any of the other slices.  Similarly, the number of detected
DF-clusters is smaller.  In calculating the total luminosity of
superclusters this additional selection effect is taken into account,
so supercluster properties are not affected.  The properties of
luminous DF-clusters of this slice are similar to the properties of
DF-clusters in other slices, and we can conclude that our 
procedure worked properly.

\subsection{Luminosities of DF-clusters in various environment}

In Paper I we used the density found with a 10~\Mpc\ smoothing as a
parameter to describe the environment in the vicinity of clusters of
galaxies.  Here we analyse the LCRS DF-clusters and loose groups to
investigate the dependence of cluster luminosities on the density of
their environment.  We calculated the global relative density $\delta$
(in units of the mean density of the low-resolution density field) for
all DF-clusters and LCRS loose groups; the results are shown in
Fig.~\ref{fig:6}.  As expected from analogy with the SDSS analysis,
there is a clear correlation between the luminosity of clusters/groups
and the density of their environment.  In all LCRS slices the relation
between the DF-cluster luminosity and the environmental density is
statistically similar.  Only in the $-6^{\circ}$ slice are low-luminosity
clusters absent due to this slice's higher luminosity completeness limit.

There exists a well-defined upper limit for the luminosity of the most
luminous clusters. DF-clusters in the highest density environments have
luminosities up to about $10^{12}~L_{\odot}$.  Most luminous loose
groups are even brighter -- their luminosity in high-density environments 
goes up to $2.5\times 10^{12}~L_{\odot}$.  The most luminous DF-clusters
in the lowest density environment have luminosities about
$10^{11}~L_{\odot}$, i.e. they are almost one-tenth as luminous.  A
similar difference was also found for the SDSS clusters.  The upper
envelope of the luminosity-density relation is statistically identical
for all LCRS slices; for the LCRS loose groups this upper envelope is
also observed, but over a smaller range of enviromental densities.

Comparing the relationship for the DF-clusters and the LCRS loose
groups shows two important differences.  First of all, there are very
few loose groups in low-density environments, $\delta \leq 0.5$ (we
recall that in this plot the environmental density is expressed in the
units of the mean density for the whole slice); there are also very
few low-luminosity groups.  This comparison shows that the LCRS loose
groups are much less suitable for studying the structure of the
universe in low-density regions.  The other difference is observed in
the regions of high environmental density.  Here the dispersion of
luminosities of loose groups is larger than that of DF-clusters.  In
other words, in high-density environments there are both high-luminosity as
well as low-luminosity loose groups, whereas most DF-clusters in high-density
environments tend to be quite luminous.  The reason for this disagreement
between the DF-clusters and the LCRS loose groups is not yet understood.

One may ask whether the cluster luminosity-density dependence could be
explained by selection effects, i.e. by the relationship between
cluster luminosities and distances shown in Fig.~\ref{fig:5}.  To
clarify this problem we divided the DF-clusters into three distance
classes and derived the luminosity-density relationship separately
for each distance class.  The results are shown in
Fig.~\ref{fig:7}.  Here the dependence of the cluster luminosity on
the density of the environment is seen quite clearly, so this effect 
must be an intrinsic property of clusters of galaxies.  Luminous
clusters are predominantly located in high-density regions, poor
clusters in low-density regions.

The luminosity-density relation can also be inverted,
telling us that we obtain a higher environmental (luminosity) density
in a given region if the DF-clusters there are more luminous.
As the environmental luminosity density comes mainly from
summing up the luminosities of individual DF-clusters, this
conclusion is trivial. Fitting a power-law density-luminosity
relationship to the data in Fig.~\ref{fig:6}, we obtain a simple
linear law, $\delta\sim L$; this means that this simplest model
may indeed be correct. Of course, this fact does not exclude other,
more complicated models of the luminosity-density dependence.

The most luminous DF-clusters in high-density environments exceed in
luminosity the most-luminous DF-clusters in low-density environments
by a factor of 10, as also found for the SDSS clusters in Paper I.
The upper envelope of the cluster luminosity-density distribution is
very well defined, as seen in Figs. \ref{fig:6} and \ref{fig:7}.  The
lower envelope is not so sharp as the upper one, and it is defined best
for nearby clusters (see the lower panels of Fig.~\ref{fig:7}).

This tendency is seen also in Fig.~\ref{fig:2}.  In the colour-coded
version of this figure (http://www.aai.ee/$\sim$maret/cosmoweb), we
see that clusters in low-density regions appear blue, which indicates
medium and small densities, whereas rich clusters, which appear red in
this figure, dominate the central high-density regions of
superclusters. This difference is very clear in nearby regions up to a
distance $\sim 300$~\Mpc.  At large distances from the observer poor
clusters cannot be observed.  Thus, at these distances, all clusters
in our appear red in our colour-coded map.

\begin{figure*}[ht]
\centering
\resizebox{0.45\textwidth}{!}{\includegraphics*{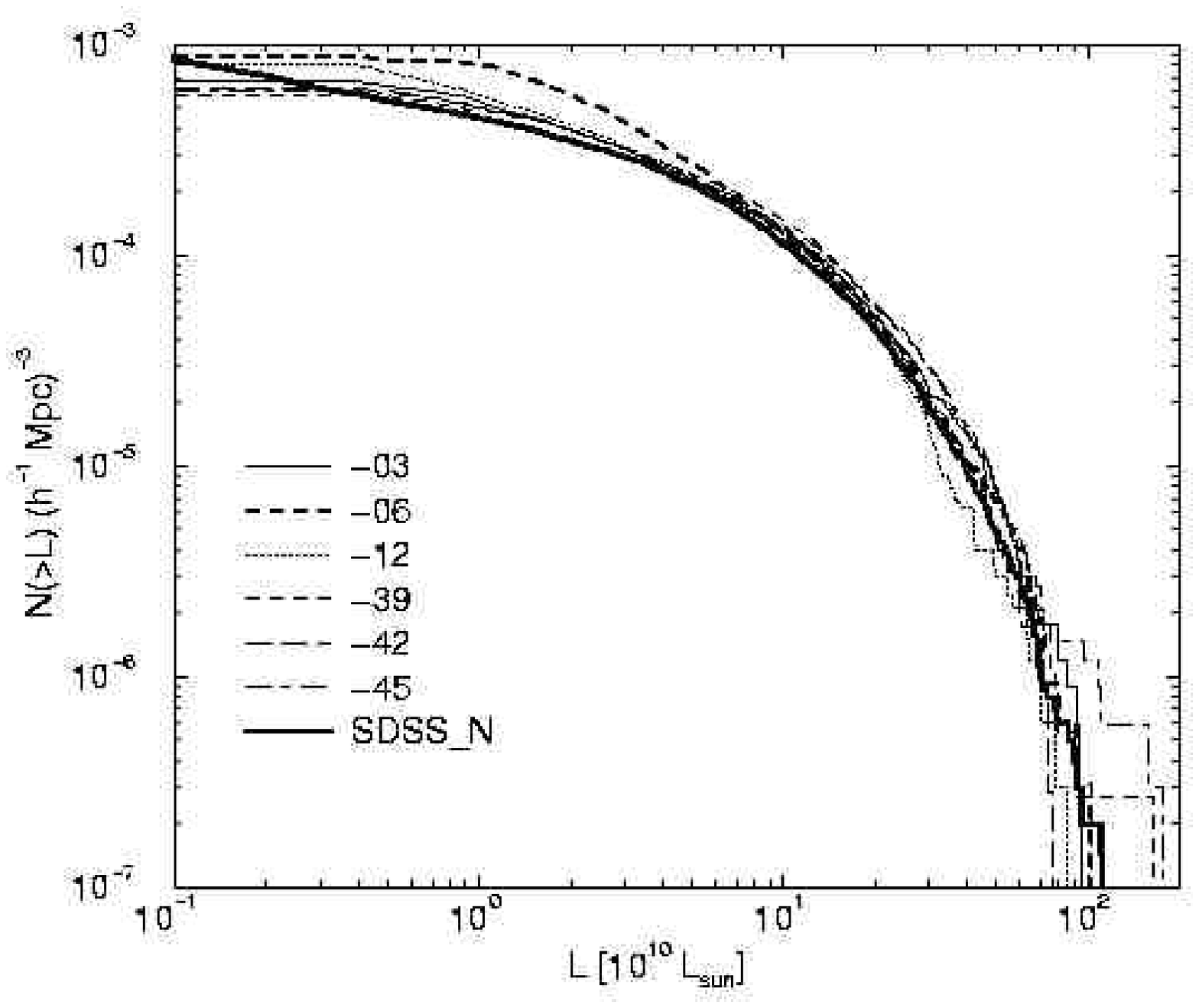}}\hspace{2mm}
\resizebox{0.45\textwidth}{!}{\includegraphics*{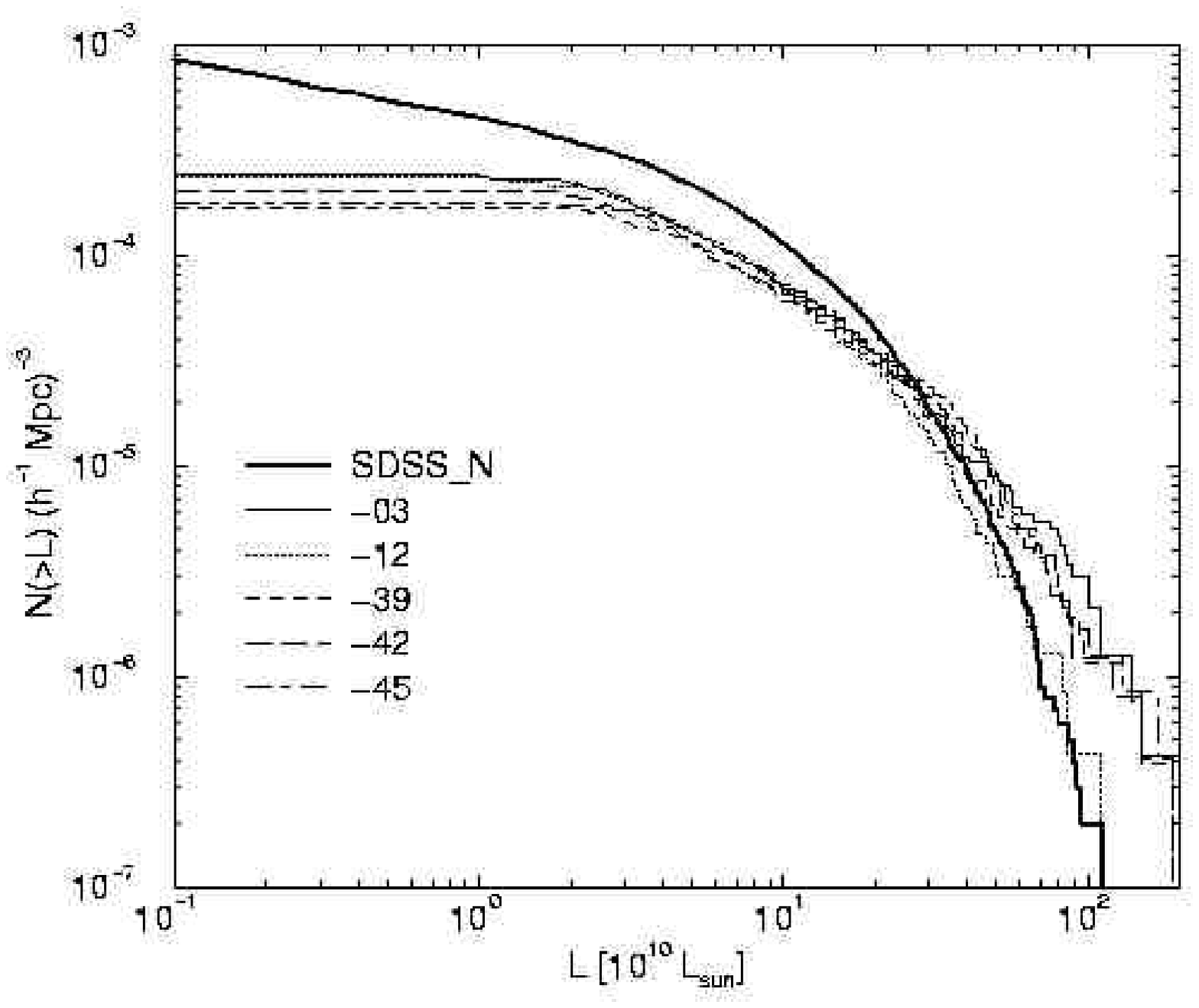}}\hspace{2mm}
\caption{The left panel shows the distribution of luminosities of
DF-clusters (the cluster luminosity function) in the LCRS slices. The
right panel shows the LCRS loose group luminosity functions.  For
comparison we show the cluster luminosity function for the SDSS
Northern slice (Paper I). }
\label{fig:8}
\end{figure*}

\subsection{The luminosity function of DF-clusters}

As in Paper I we calculated the integrated luminosity function of
DF-clusters, i.e. the number of DF-clusters per unit volume exceeding
the luminosity $L$.  As we have seen in previous sections, only the
brightest DF-clusters can be observed over the whole depth of our
samples.  We used two methods to calculate the luminosity function:
the nonparametric histogram method, and the maximum likelihood method.
In the first method we corrected for the incompleteness of less
luminous clusters by multiplying the number of observed clusters at
each luminosity step by the ratio $(d_{lim}/d_L)^3$, where
$d_{lim}=450$~\Mpc\ is the limiting distance of the total sample, and
$d_L$ is the maximum distance where DF-clusters of luminosity $L$ can
be observed.  The limiting distance for every $L$ value can be
extracted from Fig.~\ref{fig:5}; we  used here a linear relation
between $d_L$ and $\log L$.

The luminosity function for all 6 slices is shown in Fig.~\ref{fig:8}.
It spans almost 3 orders of magnitude in luminosity and 4 orders of
magnitude in spatial density.  The difference between individual
slices is very small.  Only the slice $-3^{\circ}$ has a slightly
higher density at low luminosities than the other slices. Here the
data have probably been over-corrected for non-observed poor clusters.
For comparison we plot the cluster luminosity function for the SDSS
Northern slice (Paper I). As we see there is excellent agreement
between the LCRS and the SDSS Northern slice data.

We also calculated the luminosity function of the LCRS loose groups of
galaxies; this function is shown in the right panel of
Fig.~\ref{fig:8}.  Here we used group luminosities as given by TUC.
The comparison with the DF-cluster luminosity function shows that the
luminosity of the most luminous groups is higher than in the case of
the DF-clusters (this is seen also in Figs.~\ref{fig:5} and
\ref{fig:6}).  Another difference is in the range of poor clusters.
The number of the LCRS loose groups of a given luminosity is much
lower than the number of the DF-clusters for the same luminosity. At
$L = 2 \times 10^{10}~ L_{\odot}$ the mean integrated densities of the
LCRS DF-clusters and loose groups are $4.0 \times 10^{-4}$
~(\Mpc)$^{-3}$ and $1.9 \times 10^{-4}$~(\Mpc)$^{-3}$,
respectively. For comparison we note that the densities of the SDSS
DF-clusters at the same luminosity level are $3.5 \times
10^{-4}$~(\Mpc)$^{-3}$ and $2.9\times 10^{-4}$~(\Mpc)$^{-3}$ for the
Northern and Southern slice, respectively.  The lower spatial density
of the LCRS loose groups may be explained by a selection effect
inherent in the definition of a loose group: here at least 3 galaxies
must be present in the group within the observational window, whereas
in the case of DF-clusters only one galaxy is needed. Hein\"am\"aki et al.
(\cite{hei2003}) has calculated the mass function of LCRS loose
groups.  This function also shows a lower spatial density of loose
groups in the poor cluster range.

As a second method, we describe the observed luminosity function
by the gamma-distribution, suggested by Schechter~(\cite{Schechter76}):
\begin{equation}
\Phi(L)dL = Ax^{\alpha} \exp(-x) dx,
\end{equation}
where $x=L/L^{\star}$ is the luminosity in dimensionless units, $L^{\star}$ 
is the characteristic luminosity of clusters, $A$ is the 
normalisation amplitude, and $\alpha$ is the shape parameter.  
We find the estimates of the parameters $L^{\star}$ and $\alpha$
by the maximum likelihood method (Yahil \etal \cite{yahil91}), 
minimising the log-likelihood function
\[
{\cal L}=-\sum_i^N\log(p_i),
\]
where $N$ is the number of DF-clusters and 
$p_i$ is the probability density for observing the
cluster $i$:
\[
p_i=\frac{\Phi(L_i)}{\int_{Lm(d_i)}^{Lu}\Phi(L)dL}.
\]
Here $Lu$ is the upper limit of cluster luminosities ($200 \times
10^{10}L_{\odot}$ in our case), and $Lm(d_i)$ is the lower luminosity
limit for observed clusters for the cluster distance $d_i$.  As
discussed previously, this limit is rather well defined (see
Fig.~\ref{fig:5}), although it is not easy to predict theoretically.
We defined this limit as the lower convex hull of the $d$ vs $L$
diagram.

The shape parameters for the separate slices and for the full DF-cluster
sample are given in Table~\ref{tab:DFSchechter}. The errors
are estimated by approximating the error distributions by
the appropriate $\chi^2$ distributions, as in Lin \etal\ (\cite{lin96}).
The rms errors given in the table are those for the 1-D
marginal distributions.

\begin{table}
\centering
\caption{\label{tab:DFSchechter}Shape parameters of the Schechter luminosity
function for the DF-clusters of the LCRS slices. The slices
are marked by their central declination $\delta$ (the first
column in the table). The last row gives the luminosity
function parameters for the full LCRS DF-cluster sample.}
\begin{tabular}{ccc}
\hline
$\delta (^\circ)$&$L^{\star}(10^{10}L_{\odot})$&$\alpha$\\
\hline
$-$03&15.8$\pm$1.2&-0.55$\pm$0.07\\
$-$06&15.2$\pm$1.3&-0.52$\pm$0.07\\
$-$12&14.4$\pm$1.1&-0.70$\pm$0.06\\
$-$39&16.1$\pm$1.1&-0.51$\pm$0.06\\
$-$42&13.4$\pm$0.9&-0.42$\pm$0.06\\
$-$45&19.1$\pm$1.5&-0.71$\pm$0.06\\
total&14.4$\pm$0.2&-0.43$\pm$0.01\\
\hline
\end{tabular}
\end{table}

\begin{figure}
\centering
\resizebox{.95\columnwidth}{!}{\includegraphics*{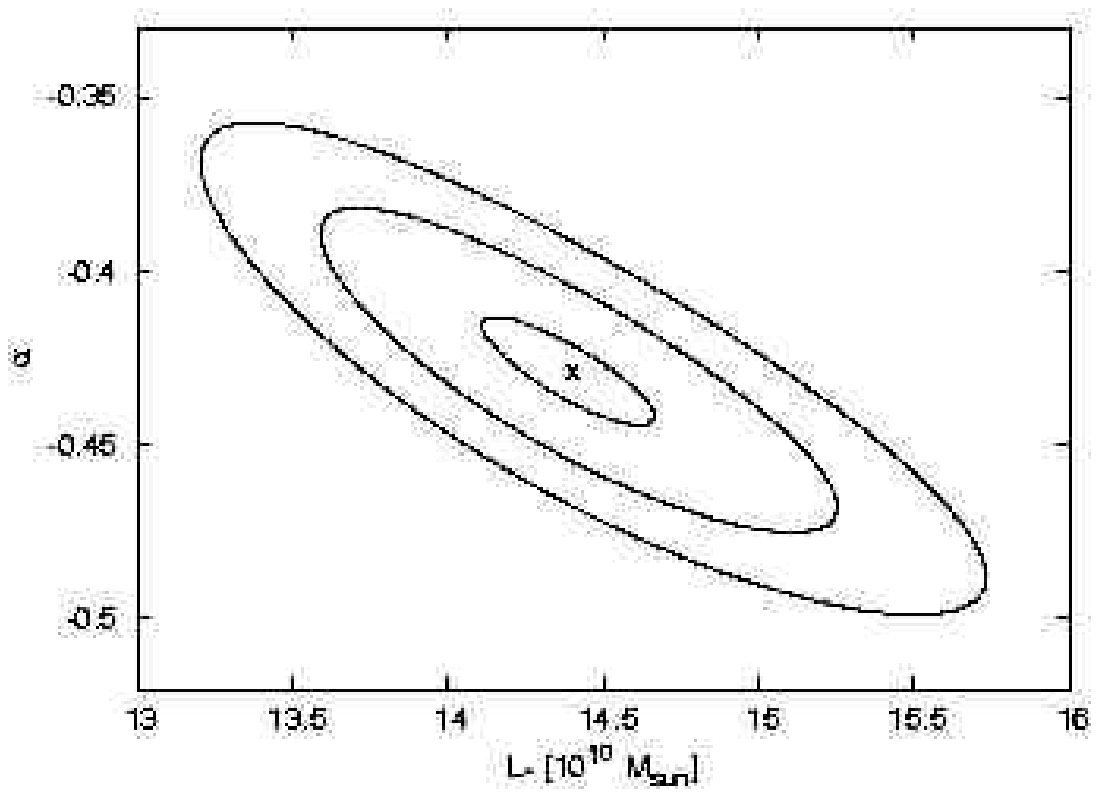}}\\
\resizebox{.95\columnwidth}{!}{\includegraphics*{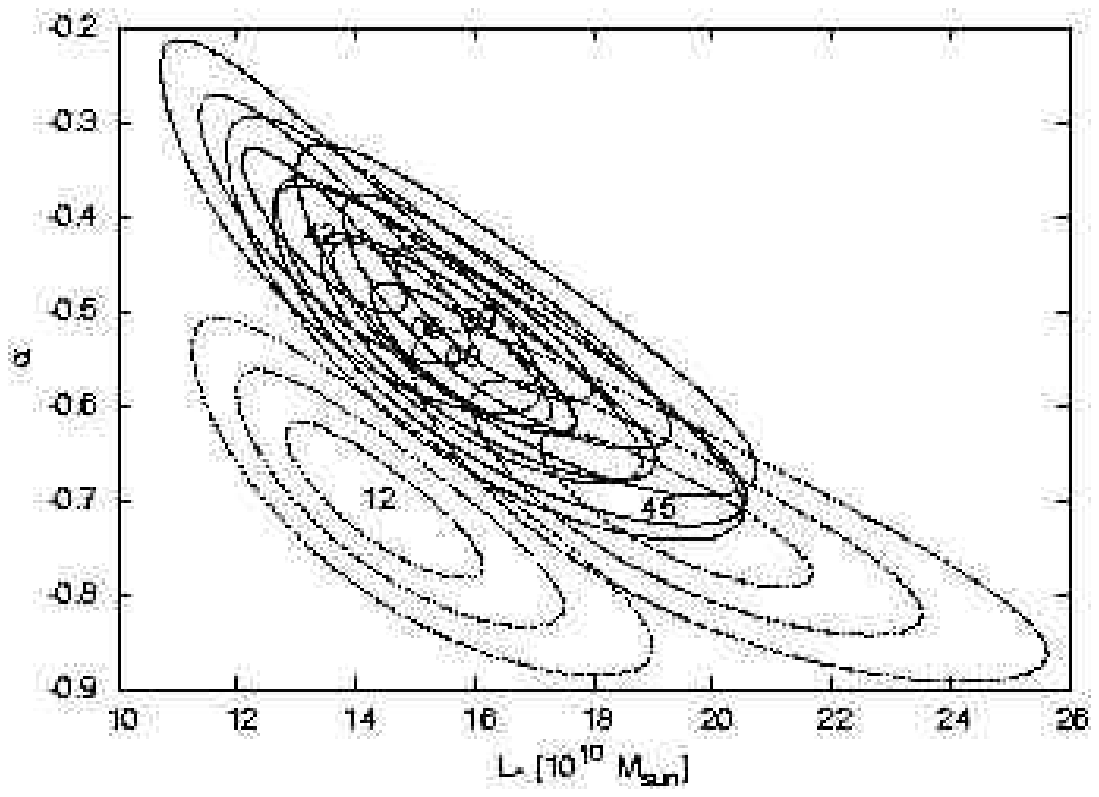}}
\caption{\label{fig:lumconf}The 1$\sigma$, 2$\sigma$, and 3$\sigma$
confidence regions for the Schechter function parameters
$L^{\star}$ and $\alpha$.
The upper panel shows the confidence regions for the total
sample and the lower panel -- the regions for individual samples.
Note that the scales in the panels are different.
The confidence regions in the lower panel are marked by
the slice number in their centres.}
\end{figure}

The 2-D 1$\sigma$, 2$\sigma$ and 3$\sigma$ (68.3\%, 95.4\%, and
99.7\%) confidence regions are shown in Fig.~\ref{fig:lumconf}.  As we
approximated the error distribution rather freely, choosing the
$\chi^2$ distribution for this purpose, these confidence levels are
approximate.  This is especially true for the confidence levels for
the outer regions, since the Schechter distribution has rather strong
wings.  The confidence regions for the total sample, shown in the
upper panel of Fig.~\ref{fig:lumconf}, are nice and narrow, but this
does not tell the whole story. The lower panel of
Fig.~\ref{fig:lumconf} shows that the confidence regions of the
parameter estimates for individual slices differ considerably.  The
slice group $-03^\circ$, $-06^\circ$, and $-39^\circ$ has similar
luminosity functions, the two slices $-42^\circ$ and $-45^\circ$ are
close to that group, but the luminosity function for the slice
$-12^\circ$ differs considerably from the rest. Estimating the rms
errors of the parameters of the luminosity function for the full
sample from the scatter of the results for the individual slices, we
find $L^* = (14 \pm 3)\times 10^{10}~L_\odot$, $\alpha = -0.44 \pm
0.15$.

To compare the LCRS DF-cluster luminosity function with that for the
SDSS slices, we also determined the Schechter parameters for these
data.  We get for the SDSS Northern slice the characteristic
luminosity $L^* = 19 \times 10^{10}~L_{\odot}$, the shape parameter
$\alpha = -0.9 $, and the amplitude $A = 4.5 \times 10^{-4}$
~(\Mpc)$^{-3}$; for the Southern slice, $L^* = 9 \times
10^{10}~L_{\odot}$, $\alpha = -0.5 $, and the amplitude $A = 10 \times
10^{-4}$ ~(\Mpc)$^{-3}$.

We discussed above that at large distances poor DF-clusters are not
visible.  This is seen in the Fig.~\ref{fig:5} luminosity vs. distance
plot, as well as in Fig.~\ref{fig:2}, where all distant clusters have
a reddish colour.  The mean luminous density is almost independent of
distance, as seen from Fig.~\ref{fig4}.  The mean constant level of
global density in the absence of poor clusters is possible only if the
luminous density due to invisible clusters (all galaxies lying outside
the visibility window) is added to luminous visible clusters.  As
discussed in Paper I, this effect makes distant clusters too luminous.
Fig.~\ref{fig:5} shows that the luminosity of the brightest
DF-clusters indeed increases with distance.  To get correct
luminosities for the DF-clusters we used in Paper I a second set of
parameters of the Schechter function to calculate weights of visible
galaxies. Here we shall use a different procedure to get correct
luminosities for the DF-clusters.

\begin{figure}[ht]
\centering
\resizebox{.95\columnwidth}{!}{\includegraphics*{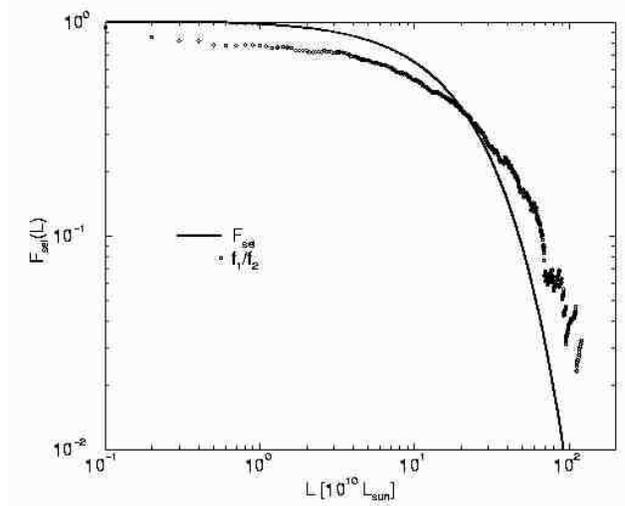}}
\caption{The cluster luminosity selection function, determined by two
  methods. The solid line shows the selection function found in this
  paper using Eq.~(\ref{eq:sel}).  Dots give the selection function as
  found in Paper I using two sets of parameters of the Schechter function.}
\label{fig:11}
\end{figure}

The fraction of the expected sum of luminosities of visible clusters to
the sum of luminosities of all clusters above a certain threshold at a
given distance from the observer can be found by
\begin{equation}
F_{sel}(L) = {\int_L^{\infty} \Phi(L) LdL \over \int_{L_0}^{\infty}
  \Phi(L) LdL},
\label{eq:sel}
\end{equation}
where $L_0= 0.5\times 10^{10} L_{\odot}$ is the lower limit of
luminosities of our cluster sample.  Using the set of Schechter
parameters for the SDSS Northern sample (which approximates well the
mean of the LCRS samples) we calculated the selection function
$F_{sel}(L)$; the results are shown in Fig.~\ref{fig:11}.  For
comparison we show also the selection function as found in Paper I
using two sets of parameters of the Schechter function, by dividing
the luminosity functions for both parameter sets at a given luminosity
$L$.  The overall agreement of the selection functions calculated by
different methods is satisfactory.  The method used in this paper is
more physically motivated.  The correction factor to calculate the
unbiased values of cluster luminosities is $1/ F_{sel}(L)$; here the
luminosity $L$ is distance dependent and should be calculated from the
lower threshold of the luminosities of the DF-clusters at a given
distance, as shown in Fig.~\ref{fig:5}. At the limiting distance
$d_{lim} = 450$~\Mpc\ the threshold luminosity is $L = 8 \times
10^{10} L_{\odot}$, and here $F_{sel} = 0.72$ (i.e. the luminosities
of DF-clusters at this distance must be decreased by a factor of 1.4).
We see that this selection effect is rather modest.

Presently we have no data for the masses of DF-clusters.  Thus we are
unable to convert the luminosity function to the cluster mass
function.  Even so, the luminosity function is interesting in and of
itself.  It is less distorted by random errors (which influence
masses of individual clusters) and it can be easily determined for all
clusters independently of the number of galaxies observed in the
cluster.  Comparison with the SDSS data shows excellent agreement.

\section{Density field superclusters}

\subsection{The DF-supercluster catalogue}

We define superclusters of galaxies as the largest non-percolating
density enhancements in the universe (Einasto et al. \cite{e1997}).
Superclusters can be identified using either galaxy or cluster data.
Here we use the low-resolution density field to find large overdensity
regions which we call density field superclusters
(DF-superclusters). This field was calculated using the galaxy data
and corrected to account for galaxies outside the visibility
window. The density field was Gaussian-smoothed, using the smoothing
length $\sigma_{sm} = 10$~\Mpc, which eliminates small-scale
irregularities and the 'finger-of-god' effect.  To reduce the conical
volume of slices (wedges) to an identical thickness we divided
densities by the thickness of the slice at the particular distance.
In this way the surface density of the field is in the mean constant.
This reduced density field for all 6 LCRS slices is shown in
Fig.~\ref{fig:3}.

\begin{figure}[ht]
\centering
\resizebox{.95\columnwidth}{!}{\includegraphics*{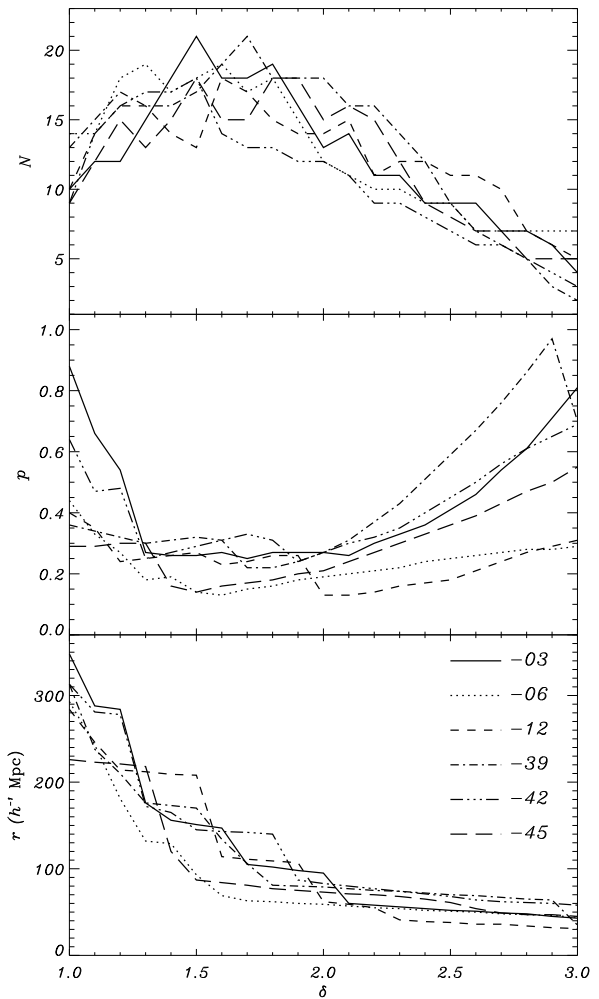}}
\caption{Properties of the LCRS density field superclusters as a
function of the threshold density, $\delta$, that separates
superclusters (high-density regions) and voids (low-density regions).
The upper panel shows the number of superclusters, $N$, the middle
panel shows the area of the largest supercluster (in units of the
total area covered by superclusters), and the lower panel shows the
size (either in the $x$ or $y$ direction, whatever is larger) of the
largest supercluster.}
\label{fig:12}
\end{figure}

{
\scriptsize
\begin{table*}[hp]
\begin{center}
\caption{The list of Northern superclusters}
\begin{tabular}{lcrrrrrrrrrrrrrl}
\hline
\\
$No$ & $\delta_{max}$&$L_{tot}$ & $L_{D}$ & $D$ & $\Delta$ & RA  & $d$ & $x$ & $y$ & $f$ & $N_{DF}$ & $N_{LC}$ & $N_{A}$ 
& Ident & Type \\ 
   &  &  &  & Mpc
 & Mpc    & deg &   Mpc   &   Mpc   &  Mpc  &     &          &     &    &    &   \\
\\ 
(1) & (2) & (3)&(4)&(5)&(6)&(7)&(8)&(9)&(10)&(11)&(12)&(13)&(14)&(15)&(16) \\
\hline 
\\ 
 
-03.01  &  3.7 &  1565 &    2595 &     31 &     39 &    156 &    184 &    106 &    151 &   0.0290 &    23 &    22 &     0 & 88 & D \\  
-03.02  &  2.1 &   161 &     379 &     14 &     18 &    156 &    114 &     65 &     94 &   0.0057 &     4 &     3 &     0 &   & D \\  
-03.03  &  2.6 &   666 &    1418 &     26 &     37 &    158 &    382 &    211 &    319 &   0.0196 &     9 &     0 &     0 &   & F \\  
-03.04  &  2.8 &   485 &     919 &     20 &     23 &    160 &    328 &    169 &    282 &   0.0119 &    13 &     0 &     0 &   & M \\  
-03.05  &  2.7 & 26798 &   20113 &     89 &    212 &    172 &    348 &    107 &    331 &   0.2306 &   124 &    29 &     4 & 100 & M \\  
-03.05a &  2.9 &   582 &    1008 &     20 &     28 &    168 &    397 &    155 &    365 &   0.0179 &    10 &     0 &     0 &   & C \\  
-03.05b &  2.7 &  8822 &    9526 &     59 &    106 &    171 &    315 &    105 &    297 &   0.1516 &    64 &    20 &     3 & 100 & D \\  
-03.05c &  4.9 &  1498 &    3695 &     34 &     43 &    173 &    430 &    130 &    410 &   0.0511 &    12 &     0 &     0 &   & M \\  
-03.05d &  3.0 &   625 &    1537 &     24 &     37 &    177 &    383 &     95 &    372 &   0.0267 &     9 &     1 &     1 & 265 & M \\  
-03.06  &  3.9 &   957 &    2480 &     30 &     36 &    183 &    329 &     42 &    326 &   0.0269 &    17 &     4 &     0 &   & M \\  
-03.07  &  2.1 &   185 &     307 &     12 &     15 &    189 &    394 &     18 &    394 &   0.0046 &     4 &     0 &     0 &   & F \\  
-03.08  &  2.4 &   527 &    1079 &     23 &     33 &    193 &    404 &    -14 &    404 &   0.0155 &     5 &     0 &     0 &   & D \\  
-03.09  &  2.2 &   268 &     523 &     16 &     22 &    196 &    420 &    -41 &    418 &   0.0077 &     5 &     0 &     0 &   & F \\  
-03.10  &  5.4 &  3266 &    5871 &     44 &     75 &    197 &    249 &    -26 &    248 &   0.0572 &    26 &     7 &     3 & 126 & M \\  
-03.11  &  2.0 &   114 &     214 &     10 &     15 &    199 &    397 &    -55 &    393 &   0.0033 &     2 &     0 &     0 &   & F \\  
-03.12  &  3.4 &  3994 &    5073 &     45 &     68 &    199 &    328 &    -41 &    326 &   0.0606 &    41 &     3 &     1 &   & M \\  
-03.13  &  2.0 &    98 &     231 &     11 &     14 &    205 &    438 &   -103 &    426 &   0.0036 &     2 &     0 &     0 &   & C \\  
-03.14  &  2.1 &   121 &     351 &     13 &     17 &    207 &    226 &    -64 &    217 &   0.0053 &     4 &     1 &     0 &   & F \\  
-03.15  &  2.0 &   173 &     303 &     12 &     16 &    208 &    432 &   -122 &    415 &   0.0046 &     5 &     0 &     0 &   & F \\  
-03.16  &  2.7 &  1979 &    3206 &     37 &     51 &    211 &    404 &   -135 &    381 &   0.0417 &    28 &     3 &     0 &   & M \\  
-03.17  &  2.2 &   210 &     575 &     17 &     25 &    216 &    240 &   -103 &    217 &   0.0084 &     6 &     2 &     0 &   & C \\  
-03.18  &  2.1 &   157 &     321 &     12 &     16 &    220 &    173 &    -84 &    151 &   0.0048 &     5 &     5 &     0 &   & F \\  
-03.19  &  2.8 &   422 &    1000 &     21 &     24 &    228 &    330 &   -197 &    264 &   0.0129 &     9 &     0 &     0 & 155 & M \\

\\ 
-06.01  &  2.2 &   197 &     398 &     18 &     17 &    154 &    427 &    242 &    352 &   0.0064 &     2 &     0 &     0 & & D \\  
-06.02  &  4.3 &  1697 &    2745 &     40 &     53 &    156 &    160 &     88 &    134 &   0.0318 &    18 &     8 &     1 & 88 & M \\  
-06.03  &  6.5 &  3911 &    6787 &     54 &     51 &    157 &    384 &    200 &    328 &   0.0596 &    23 &     2 &     2 & & M \\  
-06.04  &  1.9 &    88 &     252 &     14 &     19 &    166 &    192 &     76 &    176 &   0.0044 &     5 &     0 &     0 & & D \\  
-06.05  &  6.5 &  5728 &    7650 &     62 &     76 &    179 &    379 &     73 &    372 &   0.0774 &    32 &     3 &     2 &268 & M \\  
-06.06  &  4.6 &  3632 &    5197 &     55 &     63 &    179 &    244 &     42 &    240 &   0.0616 &    34 &    10 &     1 & & M \\  
-06.07  &  3.6 &   993 &    2027 &     36 &     35 &    190 &    401 &     -3 &    401 &   0.0257 &     6 &     0 &     0 & & F \\  
-06.08  &  2.0 &    95 &     363 &     17 &     18 &    191 &    304 &     -9 &    304 &   0.0062 &     3 &     0 &     0 & & D \\  
-06.09  &  3.4 &  1412 &    2554 &     41 &     39 &    194 &    264 &    -23 &    263 &   0.0336 &    23 &     3 &     0 & & D \\  
-06.10  &  2.6 &   624 &     984 &     27 &     29 &    202 &    439 &    -97 &    429 &   0.0144 &     7 &     0 &     0 & & F \\  
-06.11  &  2.9 &   418 &    1114 &     28 &     25 &    202 &    372 &    -86 &    362 &   0.0154 &     6 &     0 &     0 & & D \\  
-06.12  &  3.4 &  2446 &    3402 &     48 &     65 &    204 &    332 &    -85 &    321 &   0.0461 &    23 &     1 &     0 & & D \\  
-06.13  &  2.0 &   139 &     329 &     16 &     18 &    204 &    278 &    -69 &    269 &   0.0056 &     5 &     0 &     0 & & D \\  
-06.14  &  2.2 &   283 &     617 &     22 &     27 &    210 &    222 &    -79 &    208 &   0.0101 &     8 &     1 &     0 & & F \\  
-06.15  &  3.4 &  3797 &    5494 &     61 &     85 &    211 &    419 &   -157 &    389 &   0.0753 &    18 &     1 &     0 & & D \\  
-06.16  &  4.0 &  2198 &    3924 &     47 &     50 &    224 &    327 &   -185 &    270 &   0.0449 &    20 &     3 &     1 & & M \\  
-06.17  &  3.0 &   622 &    1288 &     29 &     28 &    226 &    395 &   -239 &    314 &   0.0177 &     7 &     0 &     0 & & F \\

\\ 
-12.01  &  2.9 &   711 &    2576 &     43 &     49 &    153 &    455 &    274 &    363 &   0.0352 &     8 &     0 &     0 & & D \\  
-12.02  &  2.2 &   300 &     664 &     23 &     23 &    156 &    407 &    228 &    338 &   0.0100 &     7 &     1 &     0 & & D \\  
-12.03  &  4.0 &  2077 &    3625 &     46 &     45 &    161 &    352 &    173 &    307 &   0.0400 &    26 &     4 &     0 & & D \\  
-12.04  &  2.5 &   428 &     658 &     22 &     20 &    162 &    417 &    199 &    367 &   0.0093 &     7 &     0 &     0 & & C \\  
-12.05  &  2.5 &  1412 &    2467 &     43 &     63 &    162 &    224 &    107 &    197 &   0.0354 &    15 &    11 &     0 & & M \\  
-12.06  &  3.1 & 15786 &   12738 &     91 &    123 &    173 &    341 &    104 &    325 &   0.1556 &    99 &    21 &     3 &105 & M \\  
-12.07  &  3.6 &  1312 &    2354 &     39 &     39 &    174 &    228 &     65 &    219 &   0.0282 &    20 &    10 &     0 & & M \\  
-12.08  &  3.5 &  9329 &    9187 &     73 &    107 &    197 &    271 &    -26 &    270 &   0.0994 &    64 &    26 &     2 & 118& M \\  
-12.09  &  2.4 &  6553 &    6219 &     68 &    142 &    198 &    418 &    -50 &    415 &   0.0870 &    42 &     0 &     0 & & M \\  
-12.10  &  2.7 &  2433 &    2820 &     45 &     57 &    207 &    186 &    -49 &    180 &   0.0382 &    20 &    19 &     1 &141 & M \\  
-12.11  &  2.8 &   377 &     847 &     24 &     23 &    211 &    358 &   -118 &    338 &   0.0114 &     7 &     0 &     0 & & C \\  
-12.12  &  2.1 &   383 &     602 &     22 &     28 &    211 &    299 &   -101 &    281 &   0.0093 &    11 &     0 &     0 & & F \\  
-12.13  &  2.1 &  1613 &    2627 &     43 &     61 &    222 &    314 &   -158 &    271 &   0.0353 &    21 &     4 &     1 &156 & M \\  
-12.14  &  2.8 &   517 &    1011 &     27 &     25 &    226 &    371 &   -207 &    308 &   0.0135 &     7 &     0 &     0 & & D \\  
-12.15  &  2.5 &   776 &    1418 &     33 &     33 &    227 &    427 &   -246 &    349 &   0.0202 &    17 &     0 &     1 & & D \\  
\\

\hline 
\label{tab:sc1} 
\end{tabular} 
\end{center} 
\end{table*} 

\begin{table*}[hp] 
\begin{center} 
\caption{The list of Southern superclusters} 
\begin{tabular}{rcrrrrrrrrrrrrrl} 
\\ 
\hline 
\\ 
$No$ & $\delta_{max}$& $L_{tot}$ & $L_{D}$ & $D$ & $\Delta$ & RA  & $d$ & $x$ & $y$ & $f$ & $N_{DF}$ & $N_{LC}$ & $N_{A}$ 
& Ident & Type \\ 
  &   &  &  & Mpc
 & Mpc    & deg &   Mpc   &   Mpc   &  Mpc  &     &          &
&    &   & \\
(1) & (2) & (3)&(4)&(5)&(6)&(7)&(8)&(9)&(10)&(11)&(12)&(13)&(14)&(15)&(16) \\
\hline
\\
-39.01  &  2.1 &   153 &     380 &     17 &     16 &    316 &    308 &    199 &    235 &   0.0062 &     3 &     1 &     0 & & F \\  
-39.02  &  2.1 &   202 &     303 &     15 &     14 &    316 &    263 &    170 &    200 &   0.0049 &     3 &     1 &     0 & & C \\  
-39.03  &  2.1 &   363 &     328 &     16 &     15 &    319 &    410 &    260 &    317 &   0.0053 &     6 &     0 &     0 & & C \\  
-39.04  &  2.4 &   500 &     754 &     24 &     25 &    325 &    195 &    109 &    162 &   0.0114 &     8 &     3 &     0 & & F \\  
-39.05  &  3.7 &  9818 &    9652 &     74 &     84 &    335 &    430 &    193 &    384 &   0.1086 &    40 &     0 &     1 & & M \\  
-39.06  &  2.0 &   214 &     253 &     14 &     14 &    337 &    215 &     94 &    193 &   0.0042 &     4 &     1 &     0 & & F \\  
-39.07  &  2.3 &   534 &     673 &     23 &     23 &    341 &    358 &    138 &    331 &   0.0104 &     9 &     0 &     0 & & M \\  
-39.08  &  3.2 &  2955 &    3347 &     47 &     48 &    355 &    423 &     84 &    415 &   0.0435 &    22 &     0 &     1 & & D \\  
-39.09  &  2.6 &   986 &    1498 &     33 &     35 &    356 &    300 &     58 &    295 &   0.0217 &    12 &     2 &     3 &9 & M \\  
-39.10  &  2.5 &  1740 &    2609 &     43 &     53 &      6 &    332 &     20 &    332 &   0.0373 &    16 &     4 &     2 &5 & M \\  
-39.11  &  2.1 &   607 &     766 &     25 &     37 &      7 &    199 &      8 &    199 &   0.0126 &     7 &     4 &     1 & & M \\  
-39.12  &  2.1 &   270 &     328 &     16 &     15 &      8 &    396 &     14 &    396 &   0.0054 &     4 &     0 &     0 & & D \\  
-39.13  &  2.3 &   920 &    1154 &     30 &     37 &     19 &    352 &    -37 &    350 &   0.0179 &    11 &     0 &     0 & & M \\  
-39.14  &  2.4 &   618 &     754 &     24 &     24 &     48 &    417 &   -195 &    369 &   0.0116 &     6 &     0 &     0 & & M \\  
-39.15  &  2.1 &   296 &     403 &     18 &     19 &     49 &    384 &   -179 &    340 &   0.0065 &     6 &     1 &     1 & & F \\  
-39.16  &  2.9 &  5162 &    3960 &     54 &     79 &     51 &    276 &   -136 &    240 &   0.0578 &    37 &    12 &     1 &48 & M \\  
-39.17  &  3.6 &  4702 &    4455 &     55 &     76 &     53 &    172 &    -92 &    145 &   0.0589 &    33 &    28 &     3 &48 & M \\  
-39.18  &  3.9 &  4893 &    4296 &     54 &     59 &     64 &    412 &   -253 &    326 &   0.0584 &    32 &     0 &     1 & & D \\  
\\
-42.01  &  2.0 &   131 &     241 &     14 &     13 &    321 &    310 &    179 &    254 &   0.0042 &     3 &     0 &     0 & & M \\  
-42.02  &  2.4 &   415 &     613 &     22 &     21 &    321 &    401 &    233 &    327 &   0.0098 &     8 &     0 &     1 & & M \\  
-42.03  &  2.4 &   688 &    1022 &     28 &     34 &    321 &    214 &    122 &    176 &   0.0163 &    11 &    10 &     1 &182 & F \\  
-42.04  &  3.0 &  1975 &    2715 &     43 &     46 &    336 &    332 &    136 &    303 &   0.0388 &    21 &     3 &     0 & & M \\  
-42.05  &  2.4 &   538 &     993 &     27 &     33 &    341 &    369 &    133 &    345 &   0.0156 &     7 &     0 &     1 & & F \\  
-42.06  &  3.6 &  8142 &    6669 &     66 &     93 &    353 &    269 &     60 &    263 &   0.0886 &    48 &    14 &     3 &222 & M \\  
-42.07  &  3.4 &  4840 &    4775 &     57 &     68 &      4 &    376 &     29 &    375 &   0.0657 &    31 &     3 &     1 & & M \\  
-42.08  &  2.7 &  1039 &    1437 &     32 &     32 &     20 &    356 &    -45 &    354 &   0.0213 &    17 &     0 &     0 & & M \\  
-42.09  &  1.8 &  139 &     281 &     15 &     25 &     22 &    267 &    -41 &    264 &   0.0050 &     3 &     1 &     0 & & F \\  
-42.10  &  2.0 &  200 &     328 &     16 &     20 &     30 &    365 &    -91 &    353 &   0.0056 &     3 &     1 &     1 & & F \\  
-42.11  &  3.5 &  5454 &    5404 &     57 &     75 &     47 &    188 &    -81 &    170 &   0.0680 &    36 &    21 &     1 &48 & M \\  
-42.12  &  2.0 &   180 &     369 &     17 &     18 &     50 &    427 &   -197 &    379 &   0.0064 &     3 &     0 &     0 & & F \\  
-42.13  &  2.4 &   648 &     848 &     25 &     25 &     52 &    382 &   -188 &    332 &   0.0132 &     8 &     0 &     0 & & D \\  
-42.14  &  3.4 &  1767 &    2362 &     39 &     43 &     65 &    394 &   -236 &    316 &   0.0308 &    19 &     0 &     1 & & M \\  

\\
-45.01  &  2.7 &  1193 &    1391 &     32 &     34 &    317 &    291 &    170 &    236 &   0.0203 &    15 &     2 &     2 &183 & F \\  
-45.02  &  3.2 &   822 &    1499 &     31 &     30 &    325 &    401 &    204 &    345 &   0.0199 &    10 &     0 &     2 & & D \\  
-45.03  &  2.5 &   789 &    1044 &     28 &     27 &    327 &    181 &     89 &    158 &   0.0159 &    11 &     5 &     1 &182 & D \\  
-45.04  &  2.1 &   470 &     763 &     25 &     32 &    340 &    262 &     92 &    245 &   0.0125 &     9 &     1 &     1 &197 & M \\  
-45.05  &  5.0 &  6838 &    6926 &     65 &     80 &    342 &    366 &    121 &    345 &   0.0846 &    47 &     2 &     3 &206 & M \\  
-45.06  &  2.2 &   321 &     560 &     21 &     22 &    343 &    149 &     48 &    141 &   0.0089 &     4 &     4 &     0 & & C \\  
-45.07  &  3.0 &  2220 &    2595 &     43 &     45 &      1 &    387 &     42 &    385 &   0.0366 &    24 &     3 &     0 & & D \\  
-45.08  &  2.9 &  5184 &    4863 &     59 &     71 &     16 &    365 &    -23 &    364 &   0.0687 &    39 &     4 &     1 & & M \\  
-45.09  &  2.8 &  1319 &    1577 &     33 &     35 &     20 &    434 &    -53 &    431 &   0.0226 &    13 &     0 &     0 & & D \\  
-45.10  &  3.4 &  2710 &    3635 &     48 &     52 &     24 &    261 &    -44 &    258 &   0.0471 &    23 &    14 &     0 & & M \\  
-45.11  &  2.1 &   256 &     628 &     22 &     24 &     36 &    283 &    -87 &    269 &   0.0104 &     5 &     1 &     0 & & F \\  
-45.12  &  3.1 &  1310 &    1867 &     35 &     37 &     38 &    408 &   -132 &    387 &   0.0249 &    15 &     0 &     0 & & D \\  
-45.13  &  2.7 &   556 &     885 &     25 &     24 &     46 &    384 &   -159 &    350 &   0.0129 &     6 &     0 &     1 & & F \\  
-45.14  &  2.5 &   544 &     709 &     23 &     22 &     47 &    337 &   -141 &    306 &   0.0107 &     3 &     0 &     1 & & C \\  
-45.15  &  4.7 &  6505 &    5971 &     57 &     61 &     48 &    205 &    -91 &    184 &   0.0643 &    37 &    20 &     4 &48 & M \\  
-45.16  &  2.7 &   505 &     766 &     23 &     21 &     58 &    284 &   -147 &    243 &   0.0110 &     3 &     0 &     0 & & C \\  
-45.17  &  4.8 & 11496 &    9387 &     71 &     94 &     64 &    388 &   -220 &    319 &   0.1019 &    53 &     6 &     1 & & M \\  
\\
\hline
\label{tab:sc2}
\end{tabular}
\end{center}
\end{table*}
}

In the density field approach superclusters can be identified as
connected, high-density regions.  The remaining low-density regions
can be considered voids.  To divide the density field into
superclusters and voids we need to fix the threshold density,
$\delta$, which divides the high- and low-density regions.  This
threshold density plays the same role as the neighbourhood radius used
in the friends-of-friends (FoF) method to find clusters in galaxy
samples or superclusters in cluster samples (for a more detailed
discussion see Paper I).  To make a proper choice of the threshold
density we plot in Fig.~\ref{fig:12} the number of superclusters, $N$,
the area of the largest supercluster $P$ (in units of the total area
covered by superclusters), and the maximum size of the largest
supercluster (either in the $x$ or $y$ direction), as a function of
the threshold density $\delta$ (we use relative densities as
above). The data are given for all 6 slices.  We see that the number
of superclusters has a maximum at $\delta =1.3 \dots 1.8$.  The
diameters of superclusters decrease with increasing threshold density.
At a low threshold density the largest superclusters have several
concentration centres (local density peaks), their diameters exceed
100~\Mpc, and their area forms a large fraction of the total area of
superclusters.  We have accepted the threshold density $\delta=1.8$;
the same value was also used in Paper I for the density field of the
Sloan Digital Sky Survey.  This threshold density defines compact and
rather rich superclusters.  If we want to get a sample of poor or
medium rich superclusters then we would need to use a lower threshold
density, with the price of getting supercluster complexes instead of
individual superclusters in regions of higher density.  Superclusters
were identified in the distance interval $100 \dots 450$~\Mpc.  We
include only the superclusters with areas greater than 100~(\Mpc)$^2$;
the remaining maxima are tiny spots of diameter less than 10~\Mpc.

The number of superclusters is given in Table~1.  In the
Tables~\ref{tab:sc1} and \ref{tab:sc2} we provide data on individual
superclusters; the columns are as follows: Column (1): the
identification number $No$; column (2): the peak density
$\delta_{max}$ (the peak density of the low-resolution density field,
expressed in units of the mean density); column (3): $L_{tot}$ -- the
estimated total luminosity of the supercluster, found from the sum of
observed luminosities of the DF-clusters located within the boundaries
of the supercluster; column (4): $L_D$ -- the estimated total
luminosity of the supercluster calculated by integration of the
low-resolution density field inside the boundaries of the supercluster
(both in units of $10^{10}~L_{\odot}$); column (5): $D$ -- the
diameter of the supercluster (the diameter of a circular area equal to
the area of the supercluster); column (6): $\Delta = \max(dx,dy)$ --
the maximal size of the supercluster either in the horizontal or
vertical directions (both in \Mpc); column (7): RA -- the right
ascension of the centre; columns (8) -- (10): the distance $d$ and the
coordinates, $x$, $y$, of the centre of the supercluster (in \Mpc);
column (11): $f$ -- the fraction of the area of the supercluster (in
units of the total area of superclusters in the particular slice);
columns (12) -- (14): the number of the DF-clusters $N_{DF}$, the LCRS
loose groups, $N_{LC}$, and the Abell clusters, $N_A$, within the
boundaries of the supercluster; column (15): identification with known
superclusters based on the Abell supercluster sample by E01; column
(16): the type of the supercluster, estimated by visual inspection of
the density field.

The total luminosity $L_{tot}$ was calculated as described in Paper I:
\begin{equation}
L_{tot} = {D \over D_d} L_{obs},
\end{equation}
where $L_{obs}$ is the sum of observed luminosities of DF-clusters
located within the boundaries of the DF-supercluster, $D_d$ is the
thickness of the slice at the distance of the centre of the
supercluster, and we have assumed that the size of the supercluster in
the $z-$direction coincides with its diameter in the plane of the
slice $D$.

\begin{figure}[ht]
\centering
\resizebox{.95\columnwidth}{!}{\includegraphics*{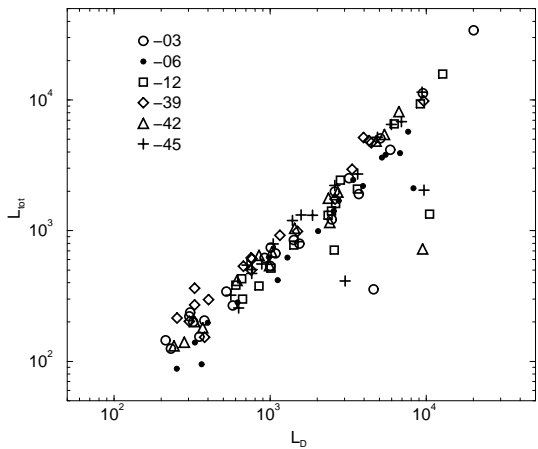}}

\caption{Total luminosities of DF-superclusters, determined by summing
  luminosities of DF-clusters, $L_{tot}$, and by integrating
  luminosity inside the threshold density contour, $L_D$.}
\label{fig:13}
\end{figure}

Comparison of the total luminosities for DF-superclusters estimated
using two different methods -- that of integrating the low-resolution
density field within the borders of the DF-supercluster ($L_D$) and
that of summing the luminosities of DF-clusters within the
DF-supercluster ($L_{tot}$) -- is shown in Fig.~\ref{fig:13}.  We see
that there are no large differences between luminosities found with
these two methods except for a few cases of distant superclusters with
a small number of DF-clusters.  We note that for the SDSS
DF-superclusters there is an even closer relationship between the
total luminosities found with the two different methods.

\subsection{Morphology of DF-superclusters}

To characterise the morphology of superclusters we estimated their
types by visual inspection of the high- and low-resolution maps.
Following Paper I we use the following classification.  If the
supercluster looks filamentary, then its type is ``F'' for a single
filament or ``M'' for a system of multiple filaments. If clusters form
a diffuse cloud and the filamentary character is not evident, then the
supercluster morphology is listed as ``D'' (diffuse); ``C'' denotes a
compact supercluster.  Tables~\ref{tab:sc1} and \ref{tab:sc2} show
that the majority of rich superclusters have a multi-filamentary
character, examples being the superclusters $-03.05$ and
$-03.10$. Compact and simple filamentary morphology is observed in
poor superclusters.

\begin{figure}[ht]
\centering
\resizebox{.95\columnwidth}{!}{\includegraphics*{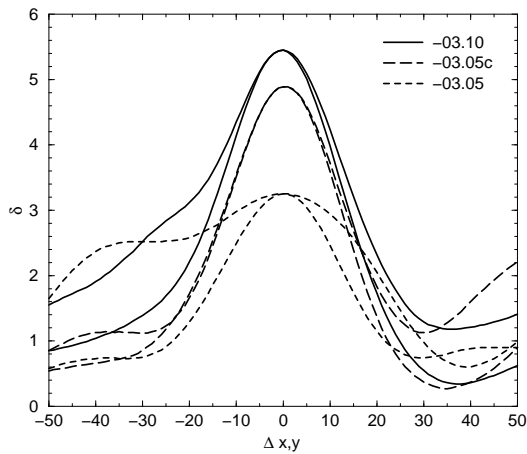}}
\caption{Characteristic density profiles in superclusters of the
  $-03^{\circ}$ slice. For each peak density profiles are given in the
  $x$ and $y$ directions. The supercluster identification is shown
  according to Table~\ref{tab:sc1}.}
\label{fig:14}
\end{figure}

The low-resolution density field map in Fig.~\ref{fig:3} shows that
low-luminosity DF-superclusters have a roundish shape, whereas high-luminosity 
superclusters have more complicated forms and contain sometimes
several concentration centres.  To see this behaviour quantitatively
we derived density profiles across the central density peak of
DF-superclusters.  Fig.~\ref{fig:14} shows several characteristic
profiles for the $-3^{\circ}$ slice.  We see that most
DF-superclusters have very symmetric density profiles.  An exception
is the largest supercluster $-03.05$ which has several concentration
centres (see the next Section), and the density peak near the
geometric centre is even lower than the peaks of one of its
sub-superclusters.

\begin{figure*}[ht]
\vspace*{7.5cm}
\caption{The total luminosities of the DF-superclusters in the LCRS slices
at different distances from the observer.} 
\includegraphics{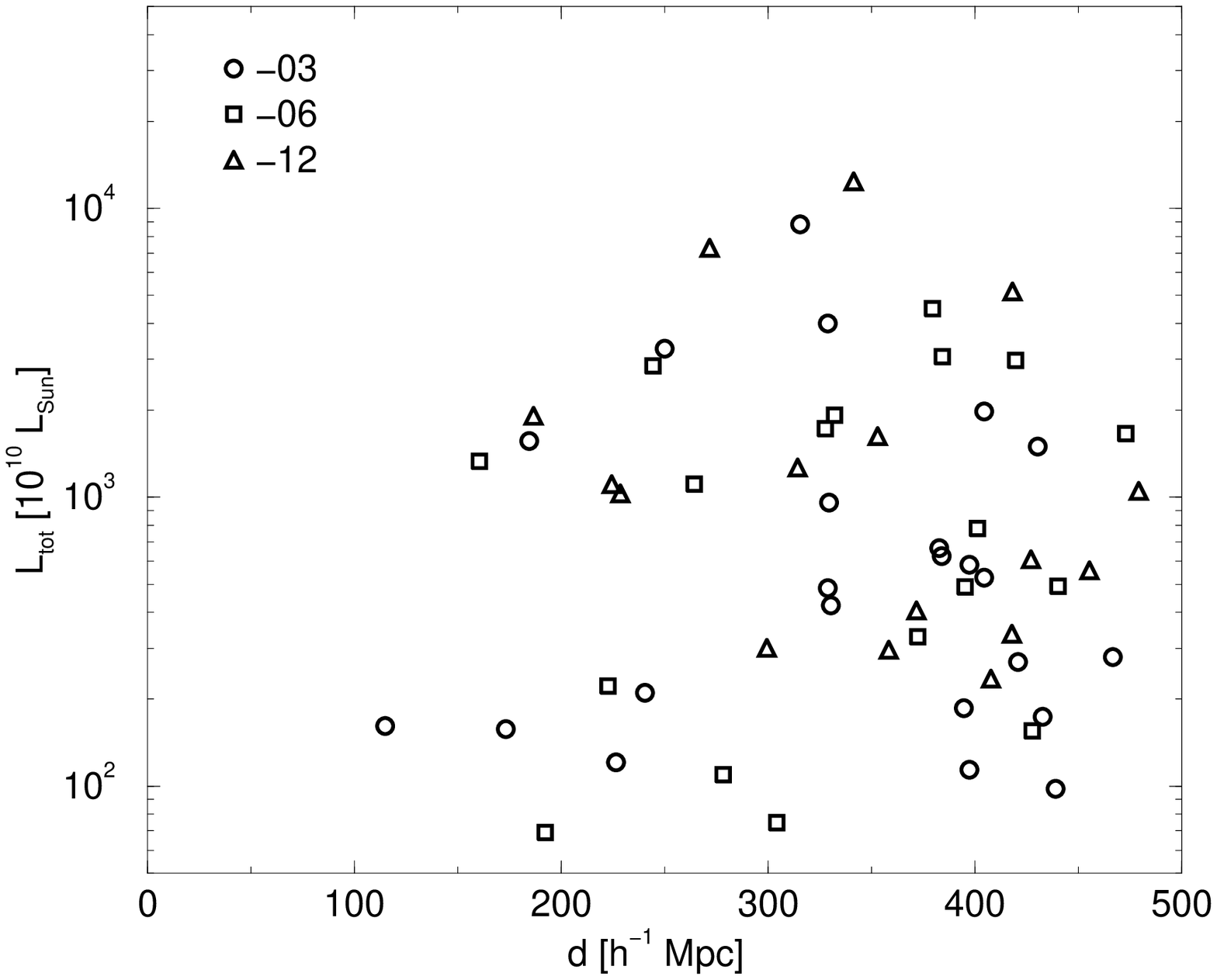}
\includegraphics{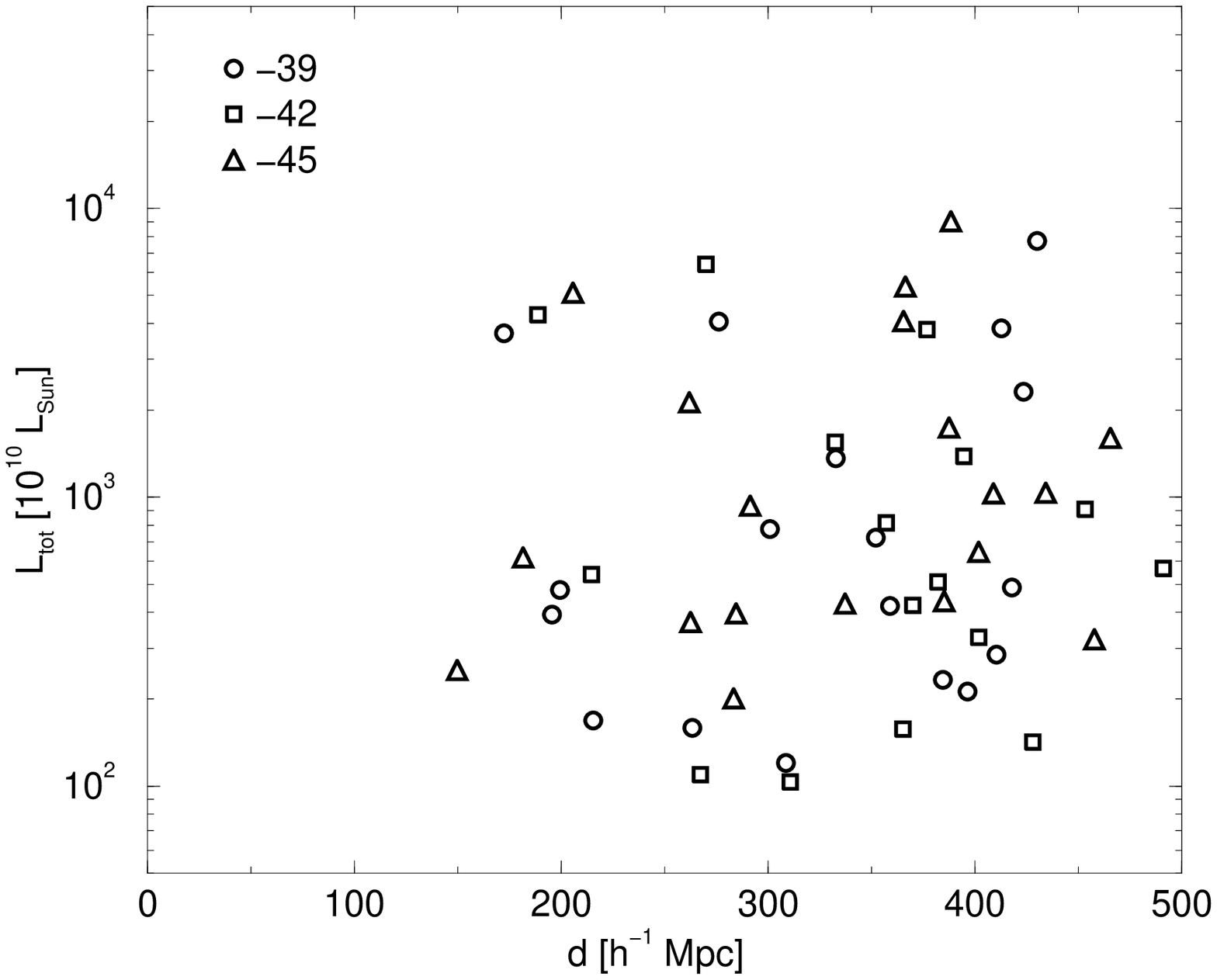}
\label{fig:15}
\end{figure*}

Fig.~\ref{fig:14} also shows that the position of the peak as the
location of the density maximum is defined rather accurately.  To
check the accuracy of the determination of the centre of
DF-superclusters we compared the positions of centres found as the
mean of extreme border coordinates in the $x$ and $y$ directions with
the positions of the peak density.  For small DF-superclusters the
difference lies within the accuracy of the determination of both
positions, $\pm 1$~\Mpc.  For large DF-superclusters with several
concentration centres the difference between various determinations of
the centre is larger (in some cases over 10~\Mpc). In Tables~\ref{tab:sc1}
and \ref{tab:sc2} we give the position of the centre as found from the mean of
the extreme coordinates.  

To check our weighting scheme we show in Fig.~\ref{fig:15} the
luminosities of the DF-superclusters as a function of the distance
from the observer, $d$.  We see that luminous DF-superclusters are
observed at various distances and that there is no obvious dependence
of supercluster luminosity on distance.  This is indirect
evidence suggesting that the luminosities of the DF-superclusters are
not influenced by large selection effects. As with the SDSS
DF-superclusters, the luminosities span an interval of over 2 orders of
magnitude.

\subsection{DF-superclusters and superclusters of Abell clusters}

Let us now discuss the structure of some prominent superclusters.  The
high-resolution map shows fine details of the structure, and the
low-resolution map shows the overall shapes and densities of the
high-density regions. The gap between adjacent slices is
rather thin, so by comparing neighbouring slices we get some
information on the 3-dimensional structure of superclusters.  Further,
the gap between the $-03^{\circ}$ slice and the Northern slice of the
SDSS survey is only about 1 degree wide, so we have a chance here to
compare the structures using both the SDSS and LCRS data.

The positions of superclusters identified from the distribution of Abell
clusters depend on a small number of objects (Abell clusters), and no
luminosity weighting is used as in the density field method.  On the
other hand, the positions of the Abell superclusters were found using
a full 3-dimensional data set, whereas the DF-superclusters were
extracted from a 2-dimensional data set. For this reason alone we cannot
expect a good coincidence in positions for the Abell and density
field superclusters.  In spite of these differences, in 19 cases the
DF-superclusters can be identified with superclusters of Abell
clusters catalogued by E01; all identifications are given in the
Tables~\ref{tab:sc1} and \ref{tab:sc2}.

The most prominent supercluster, seen both in the LCRS $-03^{\circ}$
and the SDSS Northern slices, is the SCL126 from the catalogue by E01;
in Table~\ref{tab:sc1} it is the $-03.10$; in the SDSS supercluster
catalogue the N13.  Within the $-03^{\circ}$ slice this supercluster
has 3 Abell clusters; in the SDSS survey 1 Abell cluster. These
clusters are also X-ray sources. In both slices the supercluster has a
multi-branch appearance; in the LCRS slice the filaments form a cross,
in the SDSS slice there is a strong filament in the tangential
direction (in the $y-$direction) and a weaker filament away from the
observer.  According to the calculations of the density field the
density in the region of this supercluster is one of the highest in
the whole LCRS survey. The same can be found by the distribution of
Abell clusters in this supercluster (Einasto et al. \cite{e03c}).

Another supercluster common to both the LCRS $-03^{\circ}$ and SDSS
Northern slices is the SCL155 in the catalogue by E01, the $-03.19$ in
the present catalogue, and the N23 in the SDSS catalogue (Paper
I). The main filament of this supercluster is very thin and directed
almost exactly toward the observer; individual density enhancements
can, however, be clearly distinguished.  This supercluster has also a
multi-branch appearance.

An interesting supercluster is the SCL82 (N02).  It consists of two
strong almost perpendicular filaments in the SDSS slice. In the LCRS
slice this supercluster is not visible at all.  This example shows us
that filaments in superclusters are truly thin.

The largest and most luminous supercluster in the LCRS $-03^{\circ}$
slice is the SCL100 in the Abell supercluster catalogue (the
$-03.5$ in the present catalogue). At the $1.8$ threshold density level
its length is over 200 \Mpc; at the $2.1$ level it splits into 4
sub-superclusters.  The overall form is multi-branching.  The forms of the 
sub-superclusters are different, with compact, diffuse and
multi-branch appearances.

The Sextans supercluster (SCL88 in the E01 catalogue,  $-03.01$
and $-06.02$ in the present catalogue) is clearly seen in two LCRS
slices, a weak extension (not included as a supercluster) is seen also
in the $-12^{\circ}$ slice. In the $-03^{\circ}$ slice it has a
diffuse form, but in the $-06^{\circ}$ slice it shows a clear
multi-branching character.

In the $-12^{\circ}$ slice we see two large under-dense regions
centred at $x = 20,~y = 250$
and $x = 20,~y = 350$~\Mpc, surrounded by two rings of rich
superclusters: the $-12.05, ~~-12.06, ~-12.07, ~-12.08, ~-12.09,
~-12.10, ~-12.11,~ -12.12, ~-12.13$.  Within both supervoids (we use
this term for voids surrounded by superclusters, see Lindner et
al. \cite{l95}) we see numerous small filaments of DF-clusters, but
all these clusters are poor. This example alone shows how much more
information we get using the high-resolution density field map.

The most prominent supercluster crossed by the Southern LCRS slices
(and one of the most prominent superclusters known) is the
Horologium-Reticulum supercluster (the SCL48 in E01, and the $-39.16,~
-39.17, ~-42.11, ~-45.15$ in the present catalogue).  This
supercluster contains 9 Abell clusters within the LCRS slices, 2 of
which are X-ray clusters, and a number of clusters from the APM cluster
catalogue. This supercluster has in all slices a multi-branch shape. 
In the $-39^{\circ}$ slice it is split into 2 separate superclusters.
The location of filaments in different slices is different, thus the
multi-filamentary character is seen extremely clearly.

Another very rich supercluster crossed by all Southern LCRS slices is
the $-39.18, ~-42.14, ~-45.17$. This supercluster is located at a mean
distance of 400~\Mpc\ and is too distant to be included into the E01
supercluster catalogue. In the $-45^{\circ}$ slice it consists of a
very rich DF-cluster filament, slightly inclined to the line of sight,
in the $-42^{\circ}$ slice it has also a rich DF-cluster filament,
which is directed at almost right angle in respect to the previous
one. In the $-39^{\circ}$ slice the supercluster has a diffuse shape.

Einasto et al. (\cite{e1997}) have shown that about 75\% of very rich
superclusters are concentrated in a so-called Dominant Supercluster
Plane (DSP), consisting of chains of superclusters and voids between
them.  The Southern slice $-39^{\circ}$ goes almost through the DSP,
due to this the number of Abell clusters is the largest in this slice
(28). Also the slice $\delta= -42^{\circ}$ is very close to the
DSP. The $-45^{\circ}$ slice crosses a region of extended voids
between superclusters; as elsewhere in voids this region is not
completely empty but contains numerous poor DF-cluster filaments.

\begin{table}
      \caption[]{Properties of the DF superclusters}
         \label{tab:sc}
      \[
\begin{tabular}{lrcccc}
\hline
\noalign{\smallskip}
Sample& N & $\langle N_{DF}\rangle$ 
&$\langle L_{tot}\rangle$ &$\langle L_{D}\rangle$
& $\langle f \rangle$ \\ 

(1)& (2) & (3) & (4) & (5)& (6)   \\

\hline

Abell &   24  & 34 & 5031  & 5005  & 0.05 \\
non-Abell & 44  & 12 & 1028  & 1505  & 0.02 \\

\noalign{\smallskip}
\hline
\end{tabular}
\]
{\footnotesize 

The columns  are as follows:

\noindent Column (1):
The sample type. 

\noindent Column (2): The number of superclusters in the sample.

\noindent Column (3): The mean number of DF-clusters in 
DF-superclusters.

\noindent Column (4): The mean total luminosity of
DF-superclusters, $L_{tot}$, in units of $10^{10}~L_{\odot}$ (see Tables~\ref{tab:sc1} and \ref{tab:sc2}).

\noindent Column (5): The mean total luminosity of
DF-superclusters, $L_{D}$, in units of $10^{10}~L_{\odot}$ (see Tables~\ref{tab:sc1} and\ref{tab:sc2}).

\noindent Column (6): The mean area of DF-supercluster, $f$ (in units
of the total area of superclusters in the particular slice).

}
\end{table}

Now let us compare the properties of the DF-superclusters that belong to
superclusters of Abell clusters (the Abell sample) with those of the
DF-superclusters that cannot be identified with Abell superclusters
(the non-Abell sample).  Since the data for Abell superclusters are not
as deep as the LCRS slices, we excluded all DF-superclusters more distant
than the distance limit of the catalogue of superclusters of Abell
clusters.  Table~\ref{tab:sc} shows a few properties of the Abell and
non-Abell DF-superclusters.  We see that the Abell DF-superclusters are 
about 3 times richer than the non-Abell DF-superclusters, $3 - 5$ times
more luminous, and $2 - 3$ times larger.  Most Abell DF-superclusters
have a multi-branching morphology.

\begin{figure}[ht]
\centering
\resizebox{.95\columnwidth}{!}{\includegraphics*{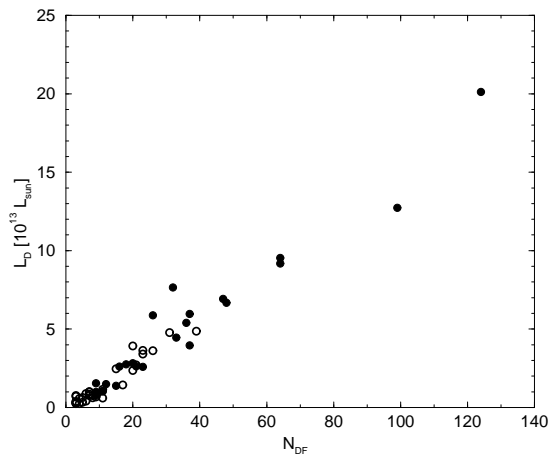}}
\caption{Total luminosities of the DF-superclusters versus the number
of DF-clusters in a supercluster (supercluster richness). Filled
circles: DF-superclusters which belong to the superclusters of Abell
clusters, empty circles: DF-superclusters that do not belong to the
Abell superclusters (see Tables \ref{tab:sc1}, \ref{tab:sc2} and text).  }
\label{fig:16}
\end{figure}

Fig.~\ref{fig:16} shows the total luminosities of superclusters versus
their richnesses (the number of DF-clusters in a supercluster).  This
figure shows that those DF-superclusters that are also the Abell
superclusters are more luminous and richer than the non-Abell
DF-superclusters.  Einasto et al. (\cite{e03a}, ~\cite{e03c}) showed
using the data on the Las Campanas loose groups (TUC) that loose
groups in superclusters of Abell clusters are richer, more
luminous, and more massive than loose groups in systems that do not
belong to Abell superclusters. Fig.~\ref{fig:16} extends this relation
to larger systems -- superclusters.  This finding shows that the
presence of rich (Abell) clusters is closely related to properties of
superclusters themselves.

\subsection{DF-clusters and superclusters, and the hierarchy of
  systems in the universe}

Abell clusters were originally identified by visual inspection of the
Palomar plates.  In spite of the subjective character of their
identification they have served for decades as the basic source of
information on high-density regions in the universe.  Now we have
redshifts and magnitudes for thousands of galaxies, which allow us to
use objective methods for cluster identification.  It is interesting
to compare the 3 sets of clusters used in this study as tracers of the
structure of the universe.

A glance at the Tables \ref{Tab1}, \ref{tab:sc1} and \ref{tab:sc2}
shows that the numbers of the DF-clusters, the LCRS loose groups, and
the Abell clusters per slice and per supercluster are very different.
Almost all Abell superclusters are seen as density enhancements in our
low-resolution density map. In contrast, there exist many
DF-superclusters and other density enhancements in the low-resolution
density field which contain no rich clusters from the Abell catalogue
within the slice boundaries.  This difference has an easy explanation:
the Abell clusters are relatively rare enhancements of the
high-resolution density field, not represented in all large-scale
density enhancements; the total number of Abell clusters within the
LCRS boundaries is about one-fiftieth the number of DF-clusters.

The sample of loose groups of galaxies by TUC contains galaxy systems
which are poorer than the Abell clusters, so the number of these
groups per DF-supercluster is much larger than the number of Abell
clusters per DF-supercluster.  However, there exist a number of
superclusters with a very small number of LCRS loose groups in it -- in
some cases there are no LCRS groups at all.  This occurs in more
distant superclusters where the LCRS groups were not searched
for. Most luminous DF-clusters can be identified with the LCRS loose
groups.  This comparison shows that among presently available cluster
samples the DF-clusters are the best tracers of structure.

Tables \ref{tab:sc1} and \ref{tab:sc2} show that in about two-thirds of cases
superclusters have a filamentary or multi-filamentary morphology. A
careful inspection of Figs. \ref{fig:2} and \ref{fig:3} indicates that
small density enhancements of the low-resolution density field have a
fine structure in the high-resolution map, similar to the
DF-superclusters. Most of these systems also consist of weak filaments
of DF-clusters in large voids. This shows the hierarchy of galaxy
systems: the morphology of galaxy systems is similar, only in
superclusters the clusters are richer, and superclusters containing
very rich clusters are themselves also richer.

\section{Conclusions}

We have used the LCRS galaxy data to construct high- and
low-resolution 2-dimensional density fields for all 6 slices of the
survey.  In calculating the density field the expected luminosity of
galaxies outside the observational window of apparent magnitudes was
estimated using the Schechter luminosity function.  The
high-resolution density field was found using a smoothing length
0.8~\Mpc, which corresponds to the characteristic scale of clusters
and groups of galaxies.  This field was used to construct a catalogue
of clusters of galaxies (DF-clusters).  The low-resolution field was
found using a smoothing length 10~\Mpc\ and was employed to construct
a catalogue of superclusters of galaxies given in Tables~\ref{tab:sc1}
and \ref{tab:sc1}.

The DF-cluster catalogue contains about 5 times more clusters/groups
than the catalogue of loose groups of galaxies compiled by TUC, and
about 50 times more than the Abell catalogue of rich clusters.  Thus,
this new sample is best suited for the investigation of the distribution
of matter in superclusters and low-density regions between
superclusters. The fine distribution of the DF-clusters in
superclusters shows that luminous superclusters preferentially have a
multi-branching structure, whereas poor superclusters as well as
galaxy systems outside superclusters have in most cases a filamentary
or compact morphology.

The density of the low-resolution field was used as an environmental
parameter to characterise the supercluster environment of the
DF-clusters.  Cluster properties depend strongly on the density of the
large-scale environment: the clusters located in high-density
environments are a factor of $5 \pm 2$ more luminous than the clusters
in low-density environments.  This finding confirms the results
obtained from the study of clusters in the Sloan Survey.

We calculated the luminosity function of the DF-clusters for all LCRS
slices, as well as for the SDSS Early Data Release samples.  These
functions can be approximated by a Schechter function with the
parameters $L^* = (14 \pm 3)\times 10^{10} L_\odot$ and $\alpha = -0.44
\pm 0.15$ (the errors are estimated from the scatter of values for
individual slices).

We found also that the DF-superclusters, which contain Abell clusters,
are more luminous and richer than the DF-superclusters without Abell
clusters.

\begin{acknowledgements}

We thank Heinz Andernach for the permission to use the new unpublished
compilation of redshifts of the Abell clusters.  The present study was
supported by the Estonian Science Foundation grants ETF 2625, ETF
4695, and by the Estonian Research and Development Council grant TO
0060058S98.  P.H. was supported by the Finnish Academy of Sciences.
J.E. thanks Fermilab and Astrophysikalisches Institut Potsdam (using
DFG-grant 436 EST 17/2/01) for hospitality where part of this study
was performed.

\end{acknowledgements}

\end{document}